\documentclass{aa}
\usepackage[varg]{txfonts}

\usepackage{graphicx}
\usepackage{caption}
\usepackage{subcaption}

\usepackage{natbib}
\bibpunct{(}{)}{;}{a}{}{,}
\usepackage[citecolor=blue, linkcolor=blue, urlcolor = black, colorlinks = true]{hyperref}

\usepackage{longtable}
\usepackage{lscape}

\usepackage{verbatim}
\usepackage{makecell}
\usepackage{multirow}
\usepackage{booktabs}
\usepackage{amsmath}
\usepackage{amssymb}

\usepackage{xspace}
\usepackage{comment}
\usepackage{textcomp}
\usepackage[separate-uncertainty=true]{siunitx}
\usepackage{xcolor} 
\usepackage[normalem]{ulem}

\newcommand{\edit}[1]{#1}

\newcommand{\gdor}{$\gamma$~Dor\xspace}
\newcommand{\Msun}{$~{\rm M_\odot}$\xspace}
\newcommand{\fov}{$f_{\rm CBM}$\xspace}
\newcommand{\mesa}{\texttt{MESA}\xspace}
\newcommand{\XcX}{$X_{\rm c}/X_{\rm ini}$\xspace}
\newcommand{\mcc}{$m_{\rm cc}$\xspace}
\newcommand{\teff}{$T_{\rm eff}$\xspace}

\newcommand{\nsample}{$\sim$14,000\xspace}
\newcommand{\percent}{~per cent\xspace}
\begin{document}

\title{Estimates of (convective core) masses, radii, and relative ages for $\sim$14,000
  {\it Gaia\/}-discovered gravity-mode pulsators monitored by TESS}
  
\author{Joey S. G. Mombarg\inst{\ref{irap},\ref{KUL},\ref{cea}}
\and Conny Aerts\inst{\ref{KUL},\ref{Radboud},\ref{MPIA}}
\and Timothy Van Reeth\inst{\ref{KUL}}  
\and Daniel Hey\inst{\ref{Hawaii}}
}
\institute{
IRAP, Universit\'e de Toulouse, CNRS, UPS, CNES, 14 avenue \'Edouard
Belin, F-31400 Toulouse,
France\\
\email{joey.mombarg@cea.fr}
\label{irap}
\and
Institute of Astronomy, KU Leuven, Celestijnenlaan 200D, B-3001
Leuven, Belgium \\
\email{conny.aerts@kuleuven.be} \label{KUL} 
\and
Universit\'e Paris-Saclay, Universit\'e de Paris, Sorbonne Paris Cit\'e, CEA, CNRS, AIM, 91191 Gif-sur-Yvette, France \label{cea}
\and Department of Astrophysics, IMAPP, Radboud University Nijmegen,
PO Box 9010, 6500 GL Nijmegen, The Netherlands\label{Radboud}
\and Max Planck Institute for Astronomy, K\"onigstuhl 17, 69117 Heidelberg, Germany\label{MPIA}
\and
Institute for Astronomy, University of Hawai'i, Honolulu, HI 96822, USA\label{Hawaii}
}

\date{Received 25 July 2024 / Accepted 4 Oct 2024}

	\abstract 
	    {Gravito-inertial asteroseismology saw its birth from the 4-year-long light curves of rotating main-sequence stars assembled by the {\it Kepler\/} space telescope. High-precision measurements of internal rotation and mixing are available for about 600 stars of intermediate mass so far that are used to challenge the state-of-the-art stellar structure and evolution models. }
	    {Our aim is to prepare for future large ensemble modelling of gravity-mode pulsators by relying on a new sample of such stars recently discovered from the third Data Release of the {\it Gaia\/} space mission and confirmed by space photometry from the TESS mission. This sample of potential asteroseismic targets is about 23 times larger than the {\it Kepler\/} sample. }
	    {We use the effective temperature and luminosity inferred from {\it Gaia\/} to deduce evolutionary masses, convective core masses, radii, and ages for $\sim$14,000 gravity-mode pulsators classified as such from their nominal TESS light curves. We do so by constructing two dedicated grids of evolutionary models for rotating stars with input physics from the asteroseismic calibrations of {\it Kepler\/} \gdor pulsators. These two grids consider the distribution of initial rotation velocities at the zero-age main sequence deduced from gravito-inertial asteroseismology, for two extreme values found for the metallicity of \gdor\ stars deduced from spectroscopy ($[{\rm M/H}] = 0.0$ and $-0.5$).}
	    {We find the new gravity-mode pulsators to cover an extended observational instability region covering masses from about 1.3\Msun to about 9\Msun. We provide their mass-luminosity and mass-radius relations, as well as convective core masses. Our results suggest that oscillations excited by the opacity mechanism occur uninterruptedly for the mass range above about 2\Msun, where stars have a radiative envelope aside from thin convection zones in their excitation layers. }
	    {Our evolutionary parameters for the sample of {\it Gaia\/}-discovered gravity-mode pulsators with confirmed modes by TESS offer a fruitful starting point for future TESS ensemble asteroseismology once a sufficient number of modes is identified in terms of the geometrical wave numbers and overtone for each of the pulsators.}
            

\titlerunning{Fundamental stellar parameters of \nsample new gravity-mode pulsators}
\authorrunning{J.S.G.\ Mombarg et al.}
\maketitle

\section{Introduction}

Data Release 3 \citep[DR3,][]{Vallenari2023} of the ESA {\it Gaia\/} space
mission \citep{Prusti2016,Brown2016,Brown2018} contains the time-series
photometry of millions of variable stars classified and delivered by
Coordination Unit\,7 \citep[CU\,7,][]{Eyer2023}.  Even if the
mission was not specifically designed for it, {\it Gaia\/}'s sparsely sampled
DR3 photometric light curves revealed excellent capacity to discover a
multitude of new nonradial pulsators. In particular,
\citet[][hereafter called Paper\,I]{DeRidder2023} discovered more than
100,000 of such pulsators occurring along the main sequence, where a
variety of excitation mechanisms are known to be active
\citep[][Chapter\,2]{Aerts2010}. 

In a follow-up study to Paper\,I, \citet[][hereafter
  Paper\,II]{Aerts2023} considered the more than 15,000 newly
discovered {\it Gaia\/} DR3 candidate gravity (g-)mode
pulsators. These turned out to have a dominant amplitude above about
4\,mmag, which was found to mark the detection threshold for nonradial
oscillations with significant frequencies occurring in the sparse DR3
light curves. The CU\,7 algorithms assigned these pulsators to either
the class of Slowly Pulsating B-type stars \citep[SPB stars
  hereafter,][]{Waelkens1991,Waelkens1998,Aerts1999,DeCatAerts2002} or
the class of $\gamma\,$Doradus stars \citep[$\gamma\,$Dor stars
  hereafter, ][]{Kaye1999,Handler1999,Uytterhoeven2011}. Both these
classes consist of multiperiodic low-frequency (typically between
0.5\,d$^{-1}$ and 3\,d$^{-1}$) g-mode pulsators with internal rotation
frequencies covering from slow to very fast compared to the
frequencies of the excited modes
\citep{Papics2017,Aerts2019-araa,GangLi2020}.

Since the era of high-cadence space photometry started, thousands of
pulsators have been discovered from their high-precision (down to
$\mu$mag) uninterrupted space photometric light curves. These are
mainly assembled with the NASA {\it Kepler\/} and Transiting Exoplanet Survey Satellite (TESS) space
telescopes \citep[see ][for an extensive review]{Kurtz2022}.  Even
though thousands of g-mode pulsators were discovered from {\it
  Kepler\/} data \citep{Blomme2010,Debosscher2011,Uytterhoeven2011},
only about 700 were characterised asteroseismically so far
(\citealt{GangLi2020,Pedersen2021}, and \citealt{Aerts2021} for a
summary), because this requires proper identification of the mode
geometry of several g~modes per star. While frequency analysis methods
from high-cadence light curves are well established
\citep[][Chapter\,5]{Aerts2010} and easy to apply
\citep[e.g. ][]{Bowman_BOOK,VanBeeck2021}, identification of the
corresponding mode geometry for each of the significant frequencies is
not a straightforward task.  For g modes, such identification can be
established from the detection of so-called period spacing patterns
revealed by modes of consecutive radial overtone, as highlighted in
the theory papers by \citet{Miglio2008,Bouabid2013} and
\citet{VanReeth2015a}. This potential was first put into practice for
two SPB stars monitored during five months by the CoRoT mission
\citep{Degroote2010,Papics2012}.

The real breakthrough in g-mode asteroseismology of main-sequence
stars came from the 4-year-long {\it Kepler\/} light curves. These
data opened up the new sub-field of gravito-inertial asteroseismology
\citep[e.g.][]{Papics2014,VanReeth2015a,VanReeth2015b,Moravveji2015,Moravveji2016,Schmid2016,VanReeth2016,Ouazzani2017,VanReeth2018,Szewczuk2018,Wu2018,Wu2019,Mombarg2019,Ouazzani2019,GangLi2020,Sekaran2021,Mombarg2021,Pedersen2021,Szewczuk2021,Szewczuk2022,Pedersen2022a,Pedersen2022b,Fritzewski2024a,Fritzewski2024b,GangLi2024}. This
led to the major conclusion that single intermediate-mass stars have
nearly rigid rotation throughout the long core-hydrogen burning phase
of their evolution, with levels of differentiality between the
near-core region and the stellar surface typically below 10\percent
\citep{GangLi2020}. The same conclusion holds for rotation frequencies
of the core \citep{Saio2021}. These findings required revisions of the
stellar evolution theory of such stars in regards to the transport of
angular momentum \citep{Aerts2019-araa,Fuller2019,Moyano2024}.

The excitation of g~modes in SPB stars is linked to the
$\kappa$-mechanism, operating around the iron opacity peak in the
partial ionisation zone of heavy elements such as iron and nickel at a
temperature of about 200\,kK. This mechanism drives an efficient
heat-engine cycle, exciting slow eigenmodes with periods of about one
day to a couple of days. In \gdor stars, the dominant excitation
mechanism is believed to be convective flux blocking. This mechanism
operates at the interface between the radiative zone and the thin
convective outer envelope, situated in an area with a temperature from
about 200\,kK to 500\,kK \citep{Guzik2000, Dupret2005}. However,
convective flux blocking alone cannot explain the observed hotter
\gdor stars, suggesting that the $\kappa$-mechanism also plays a role
in these stars \citep{Xiong2016}. The current commonly used
instability regions for the SPB stars \citep{Szewczuk2017} and \gdor
stars \citep{Dupret2005} suggest a gap between the red edge of the SPB
instability region and the blue edge of the \gdor strip, even when
considering time-dependent convection and adopting a range of
mixing-length parameters. Yet, g-mode pulsators are observed to fall
within this gap \citep[][and Paper\,I]{Balona2020,Balona2024}
indicating that at least some of the astrophysical properties of these
stars are still not fully captured by the models and/or the theory of
mode excitation.

In this paper, we estimate masses, convective core masses, radii, and
ages (in terms of the central hydrogen-mass fraction) of the largest
sample of g-mode pulsators so far. We investigate their properties
from the dominant oscillation mode discovered in the {\it Gaia\/} DR3
data and meanwhile confirmed by the TESS space photometry
\citep{HeyAerts2024}. Combining the information in these space
photometric light curves allows us to map the instability regions more
carefully than has been done so far and to decide whether there is
indeed a lack of stars with excited g~modes in the mass regime between
roughly 2 and 3\Msun.  We discuss the main astrophysical properties of
the \nsample ~{\it Gaia\/} DR3 g-mode pulsators confirmed and
reclassified as such from TESS photometry. We also provide their
potential for future asteroseismic forward and inverse modelling.

Section~\ref{sect:sample} treats the sample selection while
Sect.~\ref{sect:method} focuses on our modelling approach, which is
based on two grids of stellar models, one for solar metallicity and
one for a metallicity of ${\rm
  [M/H]}=-0.5$. Section~\ref{sect:results} covers the derived
parameters, including the mass, convective core mass, radius, and
age. We end the paper with our conclusions in
Sect.~\ref{sect:conclusions}.

\section{Sample selection} \label{sect:sample}

It was shown in Paper\,II that the {\it Gaia\/} DR3 candidate g-mode
pulsators have similar properties as the genuine SPB or $\gamma\,$Dor
pulsators observed with the {\it Kepler\/} space telescope, which
delivered 4-year-long high-cadence $\mu$mag-precision light curves
\citep{Gilliland2010,Kurtz2022}. This resemblance not only concerned
their fundamental parameters, but also the dominant frequency and its
amplitude. Nevertheless, these similarities are not a solid proof that
the candidate pulsators identified by Paper\,I are genuine class
members, because the {\it Gaia\/} DR3 data only delivered one or two
significant frequencies. Moreover, the Fourier Transforms of the {\it
  Gaia\/} DR3 light curves reveal instrumental frequencies at mmag
level, particularly for frequencies above the satellite spinning
frequency of about 4\,d$^{-1}$. Regardless whether most of the
gravito-inertial modes in SPB and \gdor stars occur below this
satellite frequency, it is difficult to exclude that the secondary
frequencies in the {\it Gaia\/} DR3 light curves deduced in Paper\,I
originate from the satellite instead of from the target stars.

In order to confirm (or refute) the nature of the {\it Gaia\/} DR3
candidate pulsators, \citet{HeyAerts2024} considered high-cadence
light curves assembled by the nominal TESS mission (having a time base
between 27\,d and 352\,d). They relied on all the candidate pulsators
found in Paper\,I having data available in the homogeneously reduced
TESS--{\it Gaia\/} light curve (TGLC) catalogue produced from
full-frame images by \citet{HanBrandt2023}. A total of 58,970 stars
out of the 106,207 {\it Gaia\/} DR3 candidate pulsators classified in
Paper\,I have TGLC light curves. Analyses of this data revealed that
more than 70\percent of the 58,970 stars have the same dominant
frequency in the completely independent TGLC and {\it Gaia\/} DR3
light curves.  Moreover, \citet{HeyAerts2024} reported that almost all
these stars turn out to be genuine multiperiodic pulsators.

A reclassification of these 58,970 TESS-confirmed pulsators was
performed by \citet{HeyAerts2024}. Here, we take all stars they
classified as \gdor/SPB or hybrid pulsator with \edit{more than two
  observed significant independent frequencies in the TESS data. This
  selection criterion is based on earlier classification work for
  highly sampled photometric light curves assembled from
  space. \citet{Debosscher2009} and \citet{Blomme2010} showed that a
  classifier based on the demand of having at least three independent
  frequencies is very effective in selecting \gdor\ pulsators from
  space photometry \citep{Tkachenko2013,VanReeth2015b}.}

\begin{figure}
    \centering
    \includegraphics[width=0.99\linewidth]{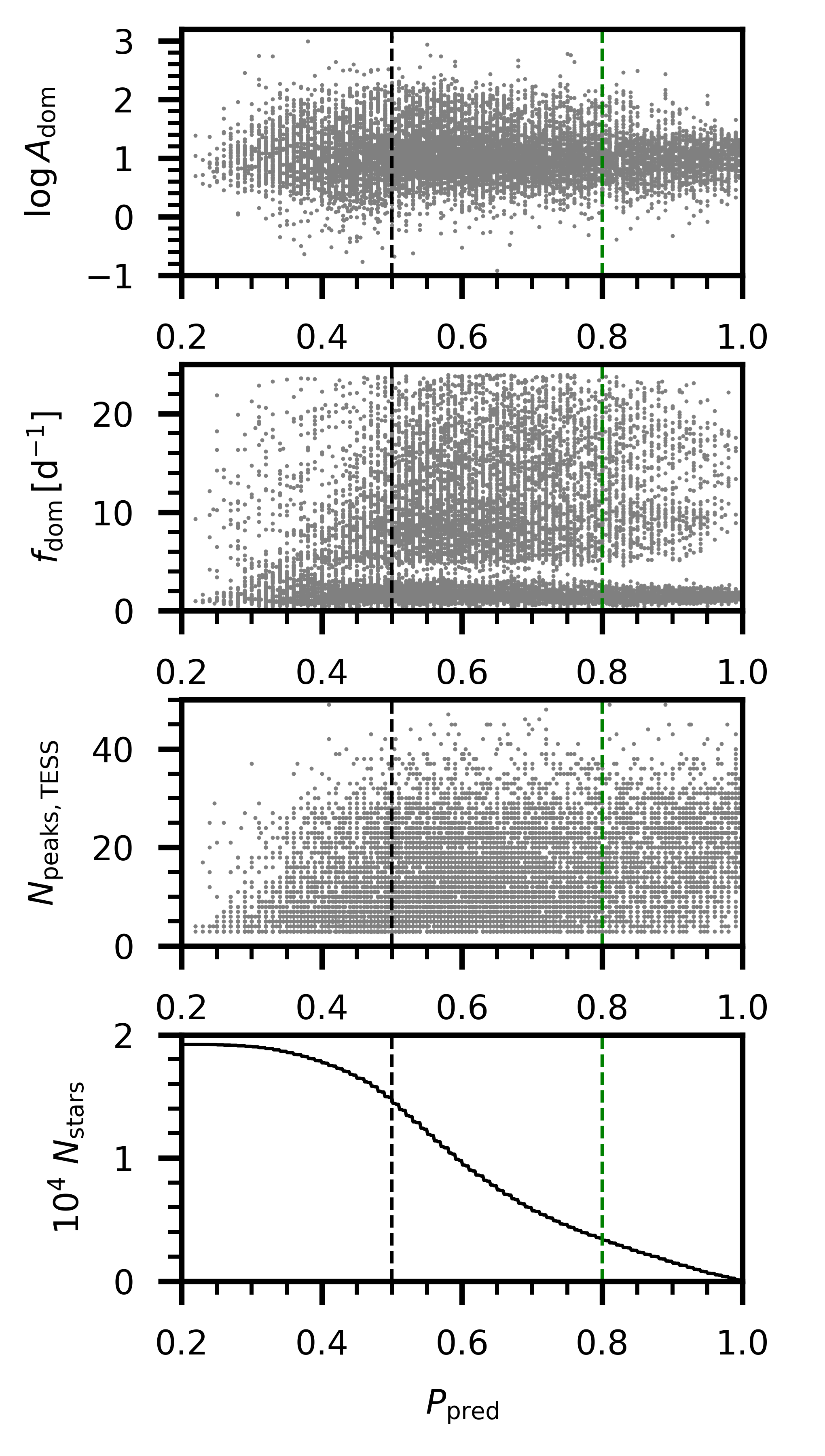}
    \caption{Pulsational characteristics of the sample. \edit{From top to bottom: amplitude of dominant
        frequency in the
        TESS light curve (in mmag), dominant frequency, and number of
        significant frequencies
        in the nominal TESS data. The bottom panel shows the number
        of stars above a given threshold on the classification
        probability $P_{\rm pred}$ from \cite{HeyAerts2024}. Vertical dashed lines indicate the used threshold in $P_{\rm pred}$ for the different samples.} }
    \label{fig:sample_char}
\end{figure}

\edit{For our study, we wish to maximise the sample size of {\it
    Gaia\/} g-mode pulsators. However, we do not want to include too
  many contaminants according to the confusion matrix of the random
  forest classifier used by \citet{HeyAerts2024}. Aside from
  demanding three significant frequencies, which greatly helps to
  avoid eclipsing binaries or rotational variables, we inspect the
  properties of the dominant frequency commonly present in the {\it
    Gaia\/} and TESS photometry.  Figure\,\ref{fig:sample_char} shows
  this frequency, its amplitude, and also the total number of
  significant frequencies in the nominal TESS light curve as a
  function of the classification probability from \citet{HeyAerts2024}.  We achieved maximal ranges of frequencies and amplitudes at $P_{\rm
    pred}\simeq 0.5$.  For this reason, we worked with the sample of
  stars having $P_{\rm pred} \geq 0.5$.}

Furthermore, we eliminated
pulsators whose effective temperature locates them outside the two
grids of stellar models discussed in Sect.~\ref{sect:method}. This
resulted in a total of \nsample new confirmed {\it Gaia\/} DR3 g-mode
pulsators, \edit{13,682 stars when we assume a metallicity of ${\rm
    [M/H] = -0.5}$, and 14,072 stars when we assume ${\rm [M/H] = 0}$,
  to be exact.} We also investigated the effect of using a lower
threshold $P_{\rm pred} \ge 0.2$ and a higher threshold $P_{\rm pred}
\ge 0.8$, on the probability of the star being a g-mode or hybrid
pulsator, which yielded a total of \edit{18,132 (18,634) and 3,165 (3,195)
  stars, respectively for ${\rm [M/H] = -0.5}$ (0.0).} These numbers
include cuts that we impose to exclude stars that fall outside of the
\edit{used grids of stellar models}, as we will discuss later.

\begin{figure*}
    \centering
    \begin{subfigure}[b]{\columnwidth}        
        \centering
        \includegraphics[width=1.06\columnwidth]{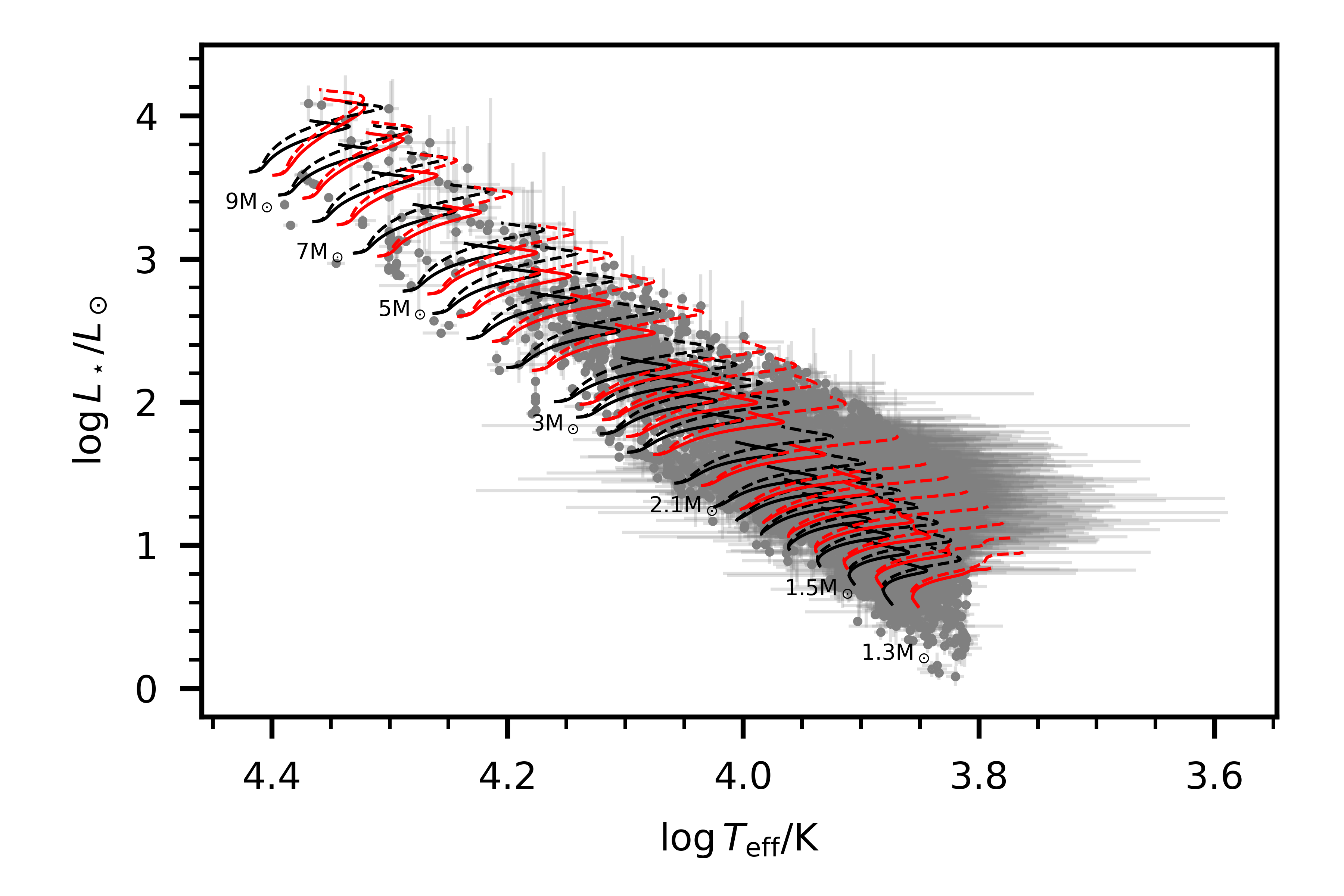}
        \caption{$Z = 0.0045$ (${\rm [M/H]} = -0.5$)}
    \end{subfigure}
    \begin{subfigure}[b]{\columnwidth}        
        \centering
        \includegraphics[width=1.06\columnwidth]{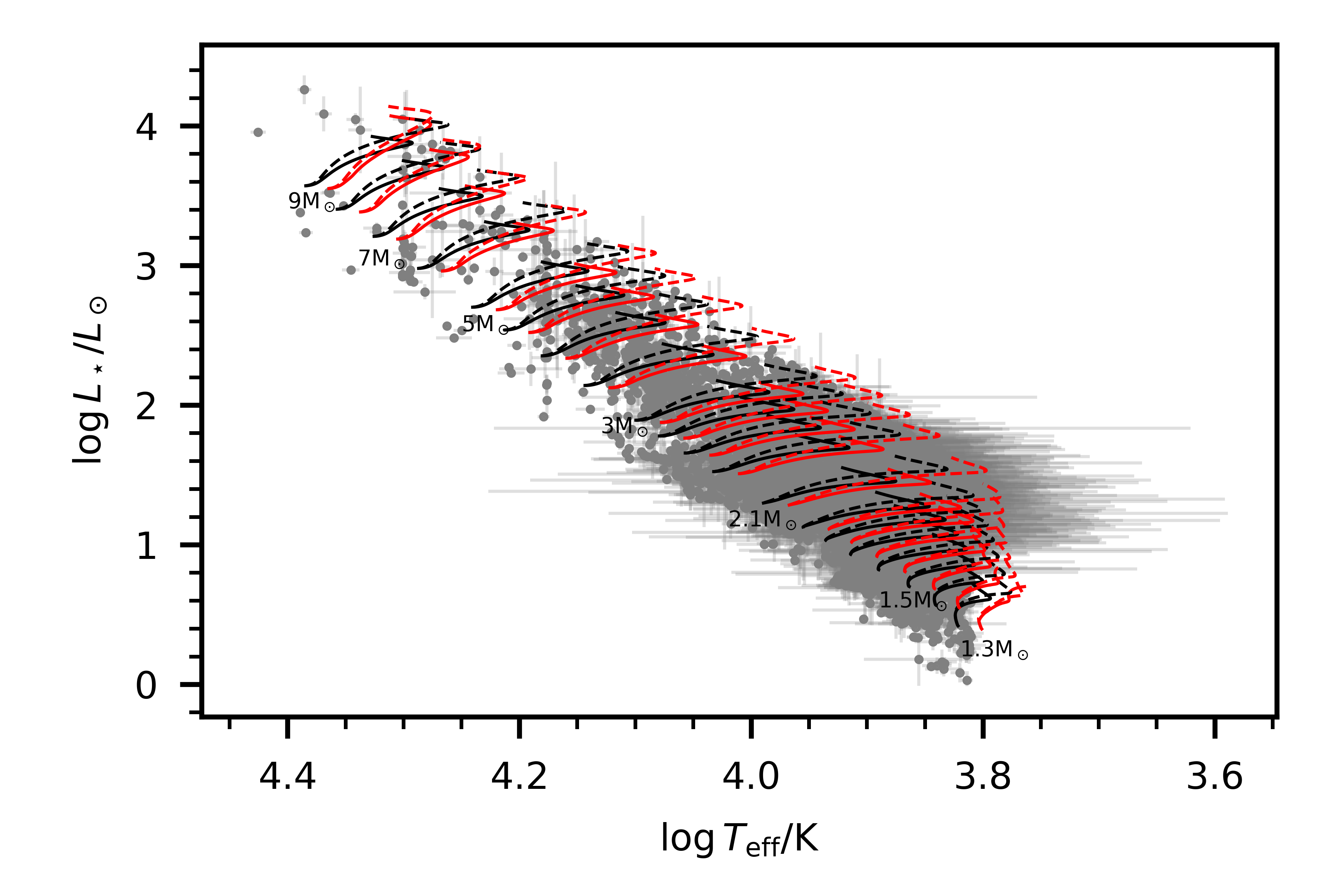}
        \caption{$Z = 0.014$ (${\rm [M/H]} = 0$)}
    \end{subfigure}
    \caption{Hertzsprung-Russell diagrams showing the sample of g-mode
      pulsators with $P_{\rm pred} \ge 0.5$ (grey symbols with error
      bars). The evolutionary tracks show the extend of the grid of
      rotating stellar models, covering a mass range of
      $[1.3,9]\,$M$_\odot$, where the black (red) colour indicates
      $\omega_0 = 0.05$ (0.55), and the solid (dashed) line style
      $f_{\rm CBM} = 0.005$ (0.025).}
        \label{fig:HRD}
\end{figure*}

The \nsample stars with $P_{\rm pred} \ge 0.5$ of being a g-mode
pulsator cover one global observational instability region as shown in
Fig.~\ref{fig:HRD}. This observational `combined {\it
  Gaia\/}-DR3/TESS' g-mode instability strip is far more extended than
the joint coverage of various theoretically predicted individual
strips for SPB and \gdor stars computed by assuming that these
pulsators are excited by convective flux blocking or the
$\kappa\,$mechanism. This was already found in Paper\,I from {\it
  Gaia\/} DR3 data and also independently concluded from a summary of
high-cadence {\it Kepler\/} and TESS space photometry by
\citet{Balona2024}. We discuss the distributions of the properties of
the pulsators, including the effect of the chosen threshold in the
classification probability on our results.

The {\it Gaia\/} DR3 sample of stars having a probability above
50\percent to be a g-mode pulsator from TESS is \edit{23} times larger
than the {\it Kepler\/} sample of 611 \gdor stars that were
asteroseismically characterised by \citet{GangLi2020}. Starting from
this {\it Kepler\/} sample, \citet{Fritzewski2024b} and
\citet{Mombarg2024} determined the mass, convective core mass, radius,
and age from grid-based modelling for subsets of respectively 490 and
539 of these pulsators, restricting to targets with identified
prograde dipole modes and their measured asymptotic period spacing
$\Pi_0$, also known as the buoyancy radius, or buoyancy travel time as
it has the dimensions of a period. On the other hand,
\citet{GangLi2020} estimated the internal rotation frequency near the
convective core from the slope of the measured period spacing patterns
from dipole prograde modes of consecutive radial order for all 611
genuine {\it Kepler\/} \gdor stars with such identified modes. For the
{\it Gaia\/} g-mode pulsators, we cannot perform asteroseismic
modelling at this stage because their modes are not identified in
terms of their geometry. This opportunity may occur in the future from
their 5-year (or longer) yearly-interrupted TESS photometry. So far,
such mode identification has only been achieved for stars in the
Continuous Viewing Zones of TESS
\citep{Garcia2022a,Garcia2022b,Fritzewski2024a,GangLi2024}.

Our aim for now is to prepare for future asteroseismic ensemble
modelling of the {\it Gaia\/}-discovered TESS-confirmed g-mode
pulsators by already deducing initial estimates for their masses,
radii, convective core masses, and ages from the homogeneously deduced
{\it Gaia\/} DR3 data. We do so for \nsample new {\it Gaia\/}/TESS
g-mode pulsators classified as such by \citet{HeyAerts2024} with a
probability above 50\percent and having the same dominant frequency in
both independent light curves. To reach our goal, we relied on their
{\it Gaia\/} DR3 effective temperature and luminosity (obtained from
the GSP-Phot analysis, Paper\,I), which are two observables available
for all stars in the sample, thus allowing us to do a homogeneous
sample analysis. To compensate for the generally underestimated
uncertainties on the effective temperature, we inflated the
uncertainties by a factor 3. This way, the typical uncertainty is more
in line with what is achieved with high-resolution spectroscopy
\citep{Gebruers2021}, \edit{namely a median uncertainty of 108\,K. The
  average relative uncertainty as a function of the effective
  temperature itself is constant, while for $\log L/{\rm L_\odot}$,
  there is a decrease from $\sim$8\percent to $\sim$3\percent between
  $1 < \log L/{\rm L_\odot} < 2$.} We coupled these observables to two
grids of stellar evolution models for rotating stars. These models
have been constructed specifically for our purpose, by relying on
input physics calibrated by the genuine {\it Kepler\/} \gdor prograde
dipole pulsators from asteroseismic modelling by
\citet{Mombarg2023}. These models are compliant with the observed
angular momentum of the stellar interior as probed by gravito-inertial
asteroseismology following the method in \citet{Aerts2018} and
summarised in \citet{Aerts2021}.

\section{Grid-based parameter estimation} \label{sect:method}

Our parameter estimation relies on the effective temperature and
luminosity from {\it Gaia\/} DR3 as the two observables. These were
coupled to the theoretically predicted values as derived from two
grids of stellar evolution models with input physics calibrated by
gravito-inertial asteroseismology. We first describe these model grids
and then the method of the parameter estimation. Our computational setup, computed models, and results discussed below are publicly available on Zenodo\footnote{\url{https://doi.org/10.5281/zenodo.13759780}}.

\subsection{Stellar structure and evolution models}
We compute a grid of stellar structure and evolution models with the
code \mesa r24.03.1 \citep{Paxton2011, Paxton2013, Paxton2015,
  Paxton2018, Paxton2019, Jermyn2023} covering masses, $M_\star$,
between 1.3 and 9\Msun. We assume an exponentially decaying
convective-boundary mixing (CBM, or overshoot) zone outside the convective core,
where the radiative temperature gradient is taken and for which the
slope of the diffusion coefficient with respect to the radial
coordinate is set by a factor \fov times the local pressure scale
height. We varied this parameter for each stellar mass between 0.005 and
0.025, based on the inferred distributions from the sample of 539
genuine {\it Kepler\/} \gdor stars by \cite{Mombarg2024}. \edit{The
  switch from convective mixing to CBM was set to a
  distance of 0.005 times the local pressure scale height into the
  convection zone (\texttt{overshoot\_f0} in \mesa).} We also varied the
initial fraction of Keplerian critical rotation at the zero-age main sequence (ZAMS),
$\omega_0$, (assuming a uniform rotation profile at the ZAMS) between
5 and 55\percent, following \citet{Mombarg2024}. In \mesa, the stellar
structure equations are modified when rotation is enabled by
redefining the radius coordinate, $r$, such that a sphere with this
radius has the same volume as the distorted surface in the Roche
model. This assumes that the angular velocity, $\Omega$, is constant
over isobars.

The chemical diffusion coefficient in the radiative envelope was
computed from the rotational mixing predicted by the theoretical works
of \cite{Zahn1992} and \cite{Chaboyer1992},
\begin{equation}
    D_{\rm rot}(r > r_{\rm cc}) = K \left( \frac{r}{N} \frac{{\rm d}\Omega}{{\rm d}r}\right)^2,
\end{equation}
where $K$ is the thermal diffusivity, $N$ the Brunt-V\"ais\"al\"a
frequency, and $r_{\rm cc}$ the radius of the convective core. The
transport of angular momentum is done in a fully diffusive
implementation, where we assumed a constant viscosity equal to
$10^7~{\rm cm^2\,s^{-1}}$. This is consistent with the range derived
for slowly rotating \gdor stars near the terminal-age main sequence
(TAMS) \citep{Mombarg2023}. The evolution of a model is stopped if
critical rotation is reached. This happens for a handful of models
only.

\edit{We note that it is currently not well known which processes
  drive chemical mixing inside the radiative envelopes of
  main-sequence stars with a convective core. This is why we mimicked
  the unknown internal mixing processes and their effect on the
  luminosity and the effective temperature by encompassing the CBM
  parameter (\fov). Furthermore, we did not include microscopic
  diffusion, but note that this type of mixing alters the local
  Rosseland mean opacity.  We compute this opacity for a solar mixture
  according to \cite{Asplund2009} using the data from the Opacity
  Project \citep{Seaton2005}. }

\edit{Adopting the present-day cosmic abundances of nearby early
  B-type stars from \citet{Nieva2012} as initial composition, we
  compute a } first grid of models for a metallicity $Z = 0.014$. This
places many of the stars below the ZAMS line (Fig.~\ref{fig:HRD},
right panel). While the error bars based on the {\it Gaia\/} data are
rather uncertain, this finding \edit{may represent shortcomings in the
  physics during the contraction phase until the ZAMS (we refer to
  \citealt{Zwintz2022} for an extensive discussion). However, the
  position of the stars below the ZAMS for solar metallicity} is also
in line with the sub-solar metallicity found consistently from both
high-resolution ground-based spectroscopy and {\it Gaia\/}
spectroscopy of \gdor\ stars obtained by \citet{Gebruers2021} and
\citet{deLaverny2024}, respectively. For this reason, we also compute
a second grid of models, for a metallicity $Z = 0.0045$ (${\rm [M/H]}
= -0.5$) found as lower limit \edit{for a large sample of
  $\gamma\,$Dor stars by \citet{deLaverny2024} from {\it Gaia\/}
  spectroscopy}. This second grid gives better agreement with the {\it
  Gaia\/} data as most stars are now well positioned above the ZAMS
line (Fig.~\ref{fig:HRD}, left panel).  \edit{For the grid with $Z =
  0.014$, an initial helium mass fraction of $Y = 0.2612$ was used, and
  for the $Z = 0.0045$ grid $Y = 0.2495$ was used, following the
  chemical enrichment rate derived by \cite{Verma2019}. The parameter
  ranges covered by the grids are listed in Table~\ref{tab:grid}. We
  limited ourselves to a lower limit of 1.3\Msun, since none of the
  g-mode pulsators known to date has a mass below this
  value. Moreover, lower-mass models require a different treatment of
  rotation as magnetic breaking becomes important.}

\edit{Convection is treated following mixing length theory
  \citep{Cox1968}, where we use the solar-calibrated value
  $\alpha_{\rm MLT} = 1.8$ \citep{Choi2016}. Since stars in the mass
  range considered here do not have a significant convective envelope
  during the main sequence, and we rely on models calibrated from
  gravity modes probing the deep interior, our work does not depend on
  the exact choice of $\alpha_{\rm MLT}$. Furthermore, an Eddington
  grey atmosphere is assumed, with an opacity that is consistent with
  the local temperature and pressure of the atmosphere.}

\begin{table}[]
    \centering
    \caption{\edit{Parameter range covered by the grids used to train the CNF.}}
    \begin{tabular}{cccc}
        \hline \hline
        Parameter & Lower boundary & Upper boundary & Step size  \\
        \hline
        $M_\star$ & 1.3\Msun & 9\Msun & 0.1-1\Msun \\ 
        \fov & 0.005 & 0.025 & 0.005 \\ 
        $\omega_0$ & 0.05 & 0.55 & 0.05 \\
        \hline
    \end{tabular}
    
    \label{tab:grid}
\end{table}

\subsection{Interpolation with conditional normalising flows} 
We seek to interpolate the computed stellar evolution models in the
three aforementioned physical quantities, $M_\star$, \fov, and
$\omega_0$. The potential of conditional normalising flows (CNFs) to
emulate grids of stellar models has been pointed out by
\citet{Hon2024}, to which we refer for details on the methodology. In
brief, normalising flows represent a class of probabilistic machine
learning tools with the capacity to make transformations between
simple and more complex distributions.  Here, we take a similar
approach as presented by \citet{Hon2024}, making use of the
\texttt{Zuko} software package \citep{Zuko}.

The general principle of normalising flows is to train a neural
network to learn the transformation for a simple probability
distribution of a random variable $z$ to a more complex distribution
of a variable $y = f(z)$. The transformation can be conditioned based
on labels $c$ such that probability densities $p(y|c)$ can be
estimated. The labels $c$ (conditional variables) are in this case
$M_\star$, \fov, and $\omega_0$. The output variables are the
effective temperature (\teff, in log scale), the luminosity
($L_\star/L_\odot$, in log scale), the rotation rate measured by the
g~modes at the convective core boundary ($\Omega_{\rm gc}$), the
hydrogen mass fraction in the core compared to the initial one (\XcX),
the mass of the convective core (\mcc/${\rm M_\odot}$), and the
stellar radius ($R_\star/R_\odot$, in log scale). The CNF is trained
on a data set of $\sim$155,000 points representing the evolution
models in each grid.

\edit{We draw 2.5 million samples from the trained CNF within the
  ranges of $M_\star$, \fov, and $\omega_0$ covered by the grid of
  stellar models.} The combination of these three conditional
variables are sampled uniformly. \edit{To avoid oversampling near the
  ZAMS and TAMS, we resample from these 2.5 million points such that
  we are left with a more or less uniform initial distribution of \XcX
  values. This leaves us with at least one million samples.} For each
star in the sample, the likelihood of each sample drawn from the CNF
is computed,

\begin{equation}
    \mathcal{L} = \prod_{y \in \bold{Y}} \frac{1}{\sqrt{2 \pi} \sigma_{y}}\exp \left( \frac{(y_{\rm CNF}- y_{\rm obs})^2}{2 \sigma_{y}^2} \right),
\end{equation}
where $\sigma_y$ denotes the observational uncertainty, and
$\bold{Y}~=~\{ \log T_{\rm eff}, \log L/L_\odot \}$. We then compute
weighted \edit{kernel density estimations (KDE) for $M_\star$, \XcX,
  \mcc/$M_\star$, and $\log R_\star/R_\odot$, where the likelihood values
  are taken as the weights. The optimal KDE bandwidth is determined
  using the Silverman method \citep{Silverman1986}.} The uncertainties
are defined as the 68 per cent confidence interval. We discard stars
that clearly fall outside the grid by imposing cuts at $3.81 < \log
T_{\rm eff} < 4.43$ and $\log L/L_\odot < 5.2 \log T_{\rm eff} - 18.3
$.

\section{Derived evolutionary stellar parameters } \label{sect:results}
\begin{figure*}
    \centering
    \includegraphics[width = 0.97\textwidth]{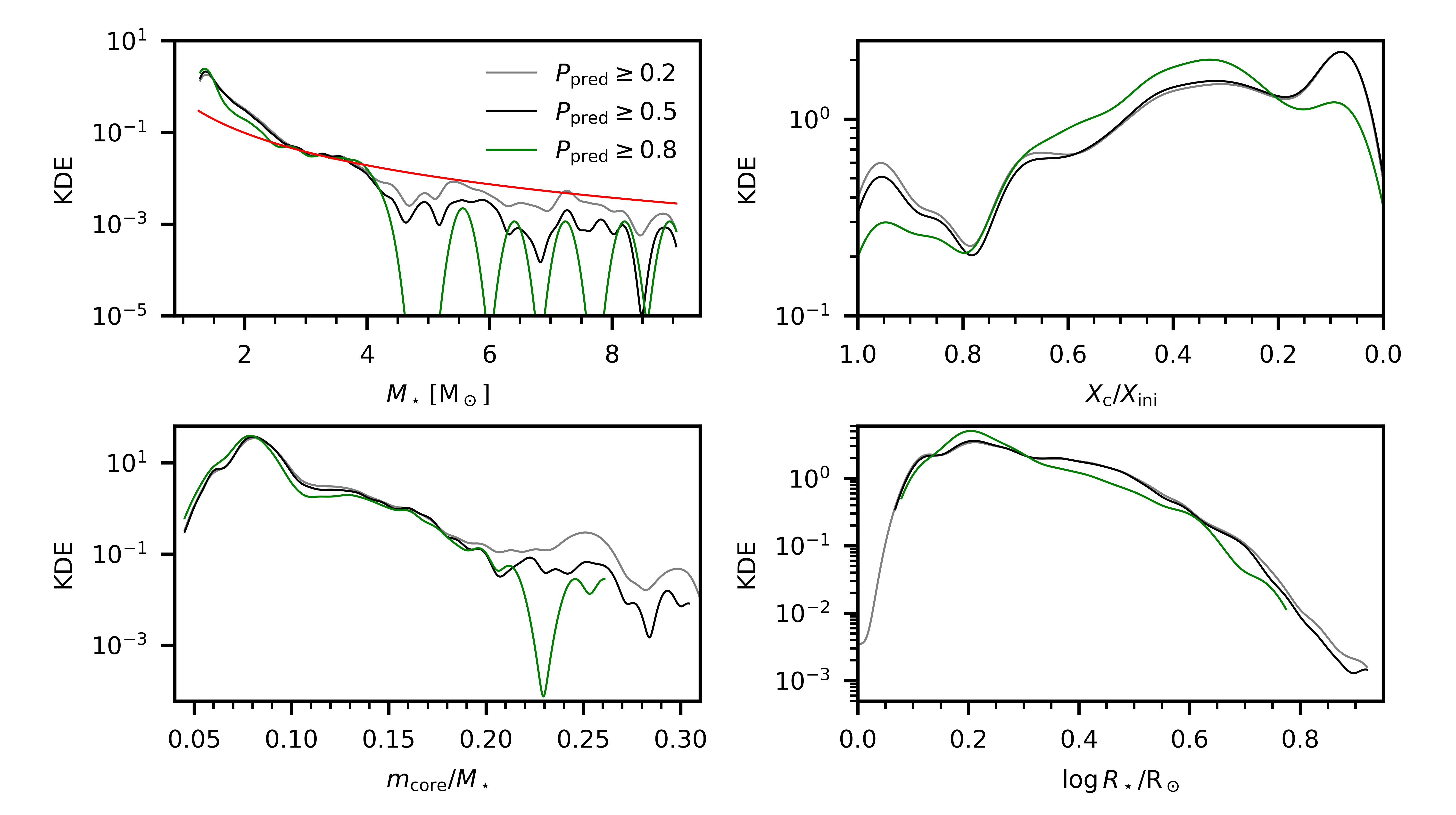}
    \caption{\edit{Kernel density estimations} of the MLEs for the
      stellar mass (top-left), the fraction of initial hydrogen in the
      core (top-right), the fractional core mass (bottom-left), and
      the stellar radius (bottom-right). The different colours
      indicate different threshold on the classification probability
      from \citet{HeyAerts2024}
      for the star to be considered a g-mode pulsator. The red line in
      the top-left panel corresponds to a Salpeter IMF with ${\rm
        d}N/{\rm d}M \propto M^{-2.35}$.}
    \label{fig:distr_P_pred}
\end{figure*}

Figure~\ref{fig:distr_P_pred} shows the resulting \edit{KDEs} of the
maximum likelihood estimates (MLEs) of the inferred masses, the
evolutionary stages (\XcX) as a proxy for the ages, the convective core
masses, and the radii. We show the results for the three different
probability thresholds of the classification results in
\citet{HeyAerts2024}. We leave out stars that fall in either the
lowest or the highest mass bin, as we consider them to be located
outside the grid. Combined with the cuts mentioned above, this omits
\edit{748 stars for $Z=0.0045$ and 373 stars for $Z=0.014$ ($P_{\rm
    pred} \geq 0.5$).}

\edit{Two-sample Kolomogorov-Smirnov tests are performed to determine
  whether a different threshold in $P_{\rm pred}$ affects the
  underlying distributions of the inferred stellar parameters. We
  adopt as null hypothesis that the underlying distributions of both
  samples are equal and reject it for p~values lower than 0.05. We
  find that the distributions for $M_\star$, \XcX, \mcc/$M_\star$ are
  not drawn for the same underlying distribution for the samples based
  on $P_{\rm pred} = 0.2$ and $P_{\rm pred} = 0.5$ (see
  Table~\ref{tab:KS_tests}). Yet, for the distribution of the radius,
  the null hypothesis cannot be rejected. When we omit stars above
  4\Msun, as higher masses are under-sampled, the null hypothesis can
  also not be rejected for the distribution of \XcX.  Furthermore,
  two-sample Kolomogorov-Smirnov tests for the distributions of
  $P_{\rm pred} = 0.5$ and $P_{\rm pred} = 0.8$ yield p~values lower
  than 0.05 for all four inferred stellar parameters. These
  conclusions hold for both $Z = 0.0045$ and $Z = 0.014$.  We do
  caution over-interpretation of these tests because the sample sizes
  are quite different (Fig.~\ref{fig:sample_char}).  }

\begin{table}[]
    \centering
    \caption{Summary of p~values from two-sample
        Kolomogorov-Smirnov tests.}
    \begin{tabular}{lllll}
        \hline \hline
        Parameter & Sample 1 & Sample 2 & $Z$ & p~value \\
        \hline
        $M_\star$ & $P_{\rm pred} = 0.2$ & $P_{\rm pred} = 0.5$ & 0.0045 &  $<0.001$  \\
        \XcX & $P_{\rm pred} = 0.2$ & $P_{\rm pred} = 0.5$ & 0.0045 &  $0.002$  \\
         &  &  &  &  0.071  \\        
        \mcc/$M_\star$ & $P_{\rm pred} = 0.2$ & $P_{\rm pred} = 0.5$ & 0.0045 &  $<0.001$  \\
        $R_\star$ & $P_{\rm pred} = 0.2$ & $P_{\rm pred} = 0.5$ & 0.0045 &  $0.477$  \\  
        \hline
         $M_\star$ & $P_{\rm pred} = 0.2$ & $P_{\rm pred} = 0.5$ & 0.014 &  $0.0011$  \\
        \XcX & $P_{\rm pred} = 0.2$ & $P_{\rm pred} = 0.5$ & 0.014 &  $0.083$  \\
         &  &  &  &  0.44  \\  
        \mcc/$M_\star$ & $P_{\rm pred} = 0.2$ & $P_{\rm pred} = 0.5$ & 0.014 &  $<0.001$  \\
        $R_\star$ & $P_{\rm pred} = 0.2$ & $P_{\rm pred} = 0.5$ & 0.014 &  $0.283$  \\  
        \hline
        $M_\star$ & $P_{\rm pred} = 0.5$ & $P_{\rm pred} = 0.8$ & 0.0045 &  $<0.001$  \\
        \XcX & $P_{\rm pred} = 0.5$ & $P_{\rm pred} = 0.8$ & 0.0045 &  $<0.001$  \\
        \mcc/$M_\star$ & $P_{\rm pred} = 0.5$ & $P_{\rm pred} = 0.8$ & 0.0045 &  $<0.001$  \\
        $R_\star$ & $P_{\rm pred} = 0.5$ & $P_{\rm pred} = 0.8$ & 0.0045 &  $<0.001$  \\ 
        \hline
         $M_\star$ & $P_{\rm pred} = 0.5$ & $P_{\rm pred} = 0.8$ & 0.014 &  $<0.001$  \\
        \XcX & $P_{\rm pred} = 0.5$ & $P_{\rm pred} = 0.8$ & 0.014 &  $<0.001$  \\
        \mcc/$M_\star$ & $P_{\rm pred} = 0.5$ & $P_{\rm pred} = 0.8$ & 0.014 &  $<0.001$  \\
        $R_\star$ & $P_{\rm pred} = 0.5$ & $P_{\rm pred} = 0.8$ & 0.014 &  $<0.001$  \\  
        \hline
    \end{tabular}
    \tablefoot{ The two additional rows for \XcX
        show the p~values when only stars below 4\Msun are taken into
        account to avoid under-representation.
        The p~value is based on the null hypothesis that
        sample 1 and sample 2 are drawn from the same distribution.}
    \label{tab:KS_tests}
\end{table}

\begin{figure*}
    \centering
    \includegraphics[width = 0.97\textwidth]{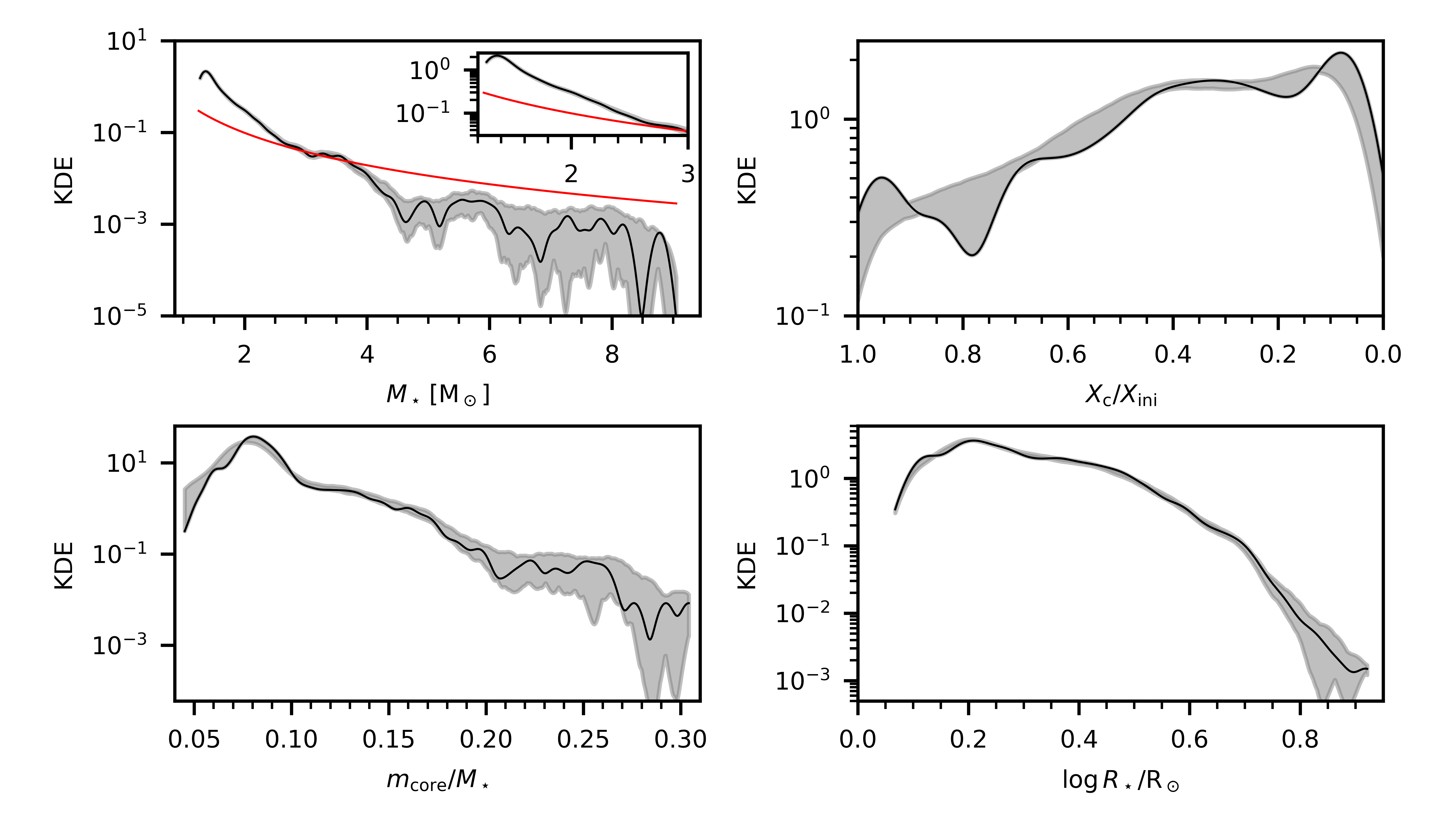}
    \caption{Same as Fig.~\ref{fig:distr_P_pred}, but for $P_{\rm
        pred} \ge 0.5$ (in black) and the result of 1000 iterations of
      bootstrapping (in grey). The inset in the top-left panel shows a
      zoom-in around the mass range of \gdor pulsators. }
    \label{fig:distr_error}
\end{figure*}

We estimate the uncertainty of the derived distributions by means of
bootstrapping, where each measurement is perturbed
\edit{independently} within its uncertainty interval following a
uniform probability\footnote{Note that the resulting distributions of
the MLEs do not necessarily follow a normal distribution. Therefore,
we take the more conservative assumption to sample from a uniform
distribution within the 68\%-confidence interval. }. The results of
1000 iterations are shown in grey in Fig.~\ref{fig:distr_error}. We
conclude that the shape of the derived distributions is robust,
although the uncertainties at higher masses are larger as a result of
the scarcity of higher-mass g-mode pulsators.  \edit{The derived
  distribution of the stellar masses is compared with a Salpeter
  initial-mass function (IMF, \citealt{Salpeter1955}) ${\rm d}N/{\rm
    d}M \propto M^{-2.35}$. We observe an excess of g-mode pulsators
  below roughly 1.6\Msun.  This suggests that the convective flux
  blocking mechanism is efficient in exciting g~modes within a narrow
  mass range.}

\begin{figure}
    \centering
    \includegraphics[width=\columnwidth]{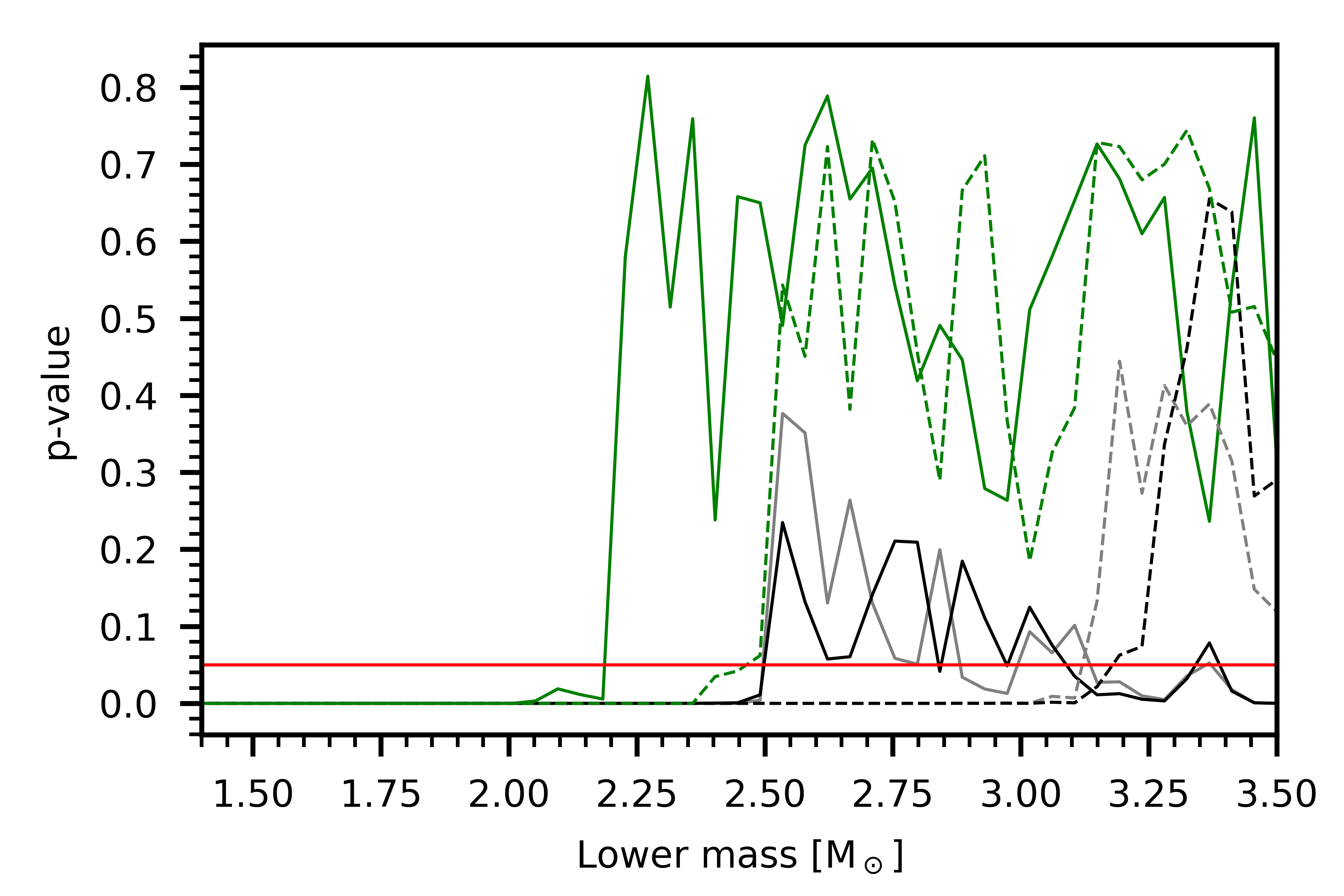}
    \caption{Resulting p~value from a one-sample Kolomogorov-Smirnov
      test between the inferred mass distribution of the sample of
      g-mode pulsators and a Salpeter IMF, as a function of the mass
      range tested. The upper mass is always 4\Msun. The solid lines
      correspond to $Z = 0.0045$, the dashed lines correspond to $Z =
      0.014$. The solid red line indicates a p~value of 0.05 below
      which we reject the null hypothesis. }
    \label{fig:KS_IMF}
    \end{figure}

Interestingly, we do not observe a drop in the number of g-mode
pulsators between the predicted blue edge of the \gdor instability
region \citep[roughly above 2\Msun,][]{Dupret2005} and the predicted
red edge of the SPB instability region \citep[roughly below
  3\Msun,][]{Szewczuk2017}.  \edit{We perform one-sample
  Kolomogorov-Smirnov tests assuming a cumulative distribution
  function predicted by a Salpeter IMF. We adopt as null hypothesis
  that the sample is drawn from the assumed IMF distribution. Again,
  we omit stars with a mass above 4\,M$_\odot$ to avoid
  under-representation. When using a variable lower limit on the
  stellar mass and a fixed upper limit of 4\Msun, we find that the
  mass distribution of our sample of g-mode pulsators is compatible
  with a Salpeter IMF down to roughly 2.2-2.5\Msun, even for the
  strict threshold of
  $P_{\rm pred} \geq 0.8$, as shown in Fig.~\ref{fig:KS_IMF}. The
  resulting deduced mass distribution of the {\it Gaia\/} pulsators
  therefore indicates that the $\kappa$-mechanism is effective in
  exciting g~modes between the currently assumed disjoint theoretical
  instability regions of $\gamma\,$Dor and SPB stars.}
  
  {As already highlighted in the Introduction, observational evidence for pulsators in this part of the 
 Hertzsprung-Russell diagram (HRD) had already been found from high-cadence space photometry assembled with the {\it Kepler\/} and TESS space telescopes,
 prior to the results from {\it Gaia\/} DR3 by \citet{DeRidder2023} and \citet{Aerts2023}, and \citet{HeyAerts2024}. Even earlier  \citet{Mowlavi2013,Mowlavi2016} detected fast rotating g-mode pulsators with dominant amplitudes of a few mmag in that part of the HRD of a young open cluster, while \citet{Degroote2009a} detected many candidate pulsators in that region from CoRoT space photometry as well. Along with the summary by \citet{Balona2024} it seems firmly established now that g-mode pulsations occur all along the main sequence. 
  Our results on the mass distribution are in line with these earlier observational studies.}

\begin{figure}
    \centering
    \includegraphics[width = \columnwidth]{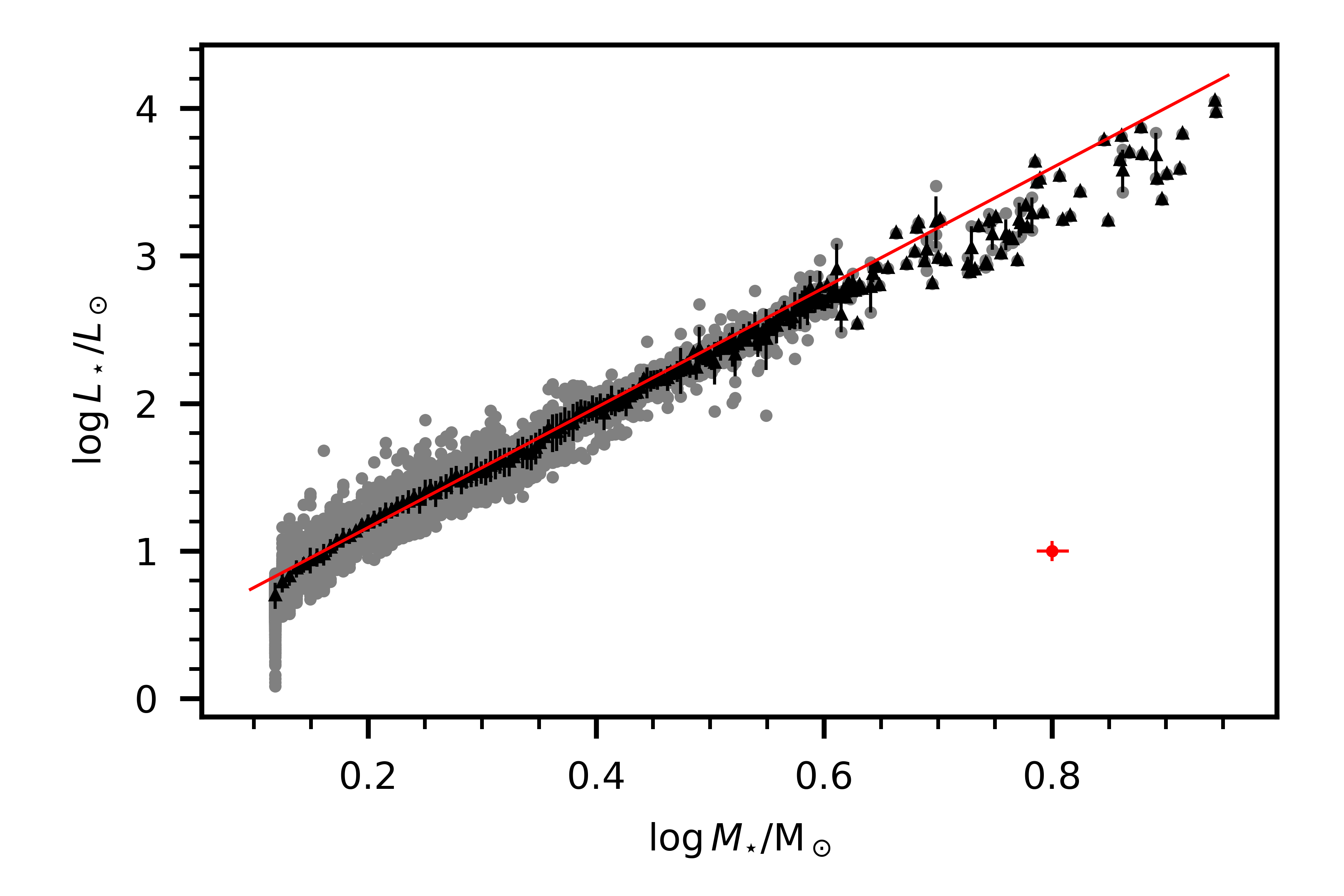}
    \caption{Relation between the inferred stellar mass and the
      measured luminosity from Gaia of the stars in the
      sample. \edit{The solid red line indicates the fitted
        mass-luminosity relation, yielding $\log({L_\star}/{L_\odot})
        = {4.0623(6)} \log\left ({M_\star}/{M_\odot} \right) +
        0.3468(2)$. The black triangles indicate bin averaged-values.}
      Average uncertainties are indicated by the red data point.}
    \label{fig:mass_vs_lum}
\end{figure}

\begin{figure}
    \centering
    \includegraphics[width = \columnwidth]{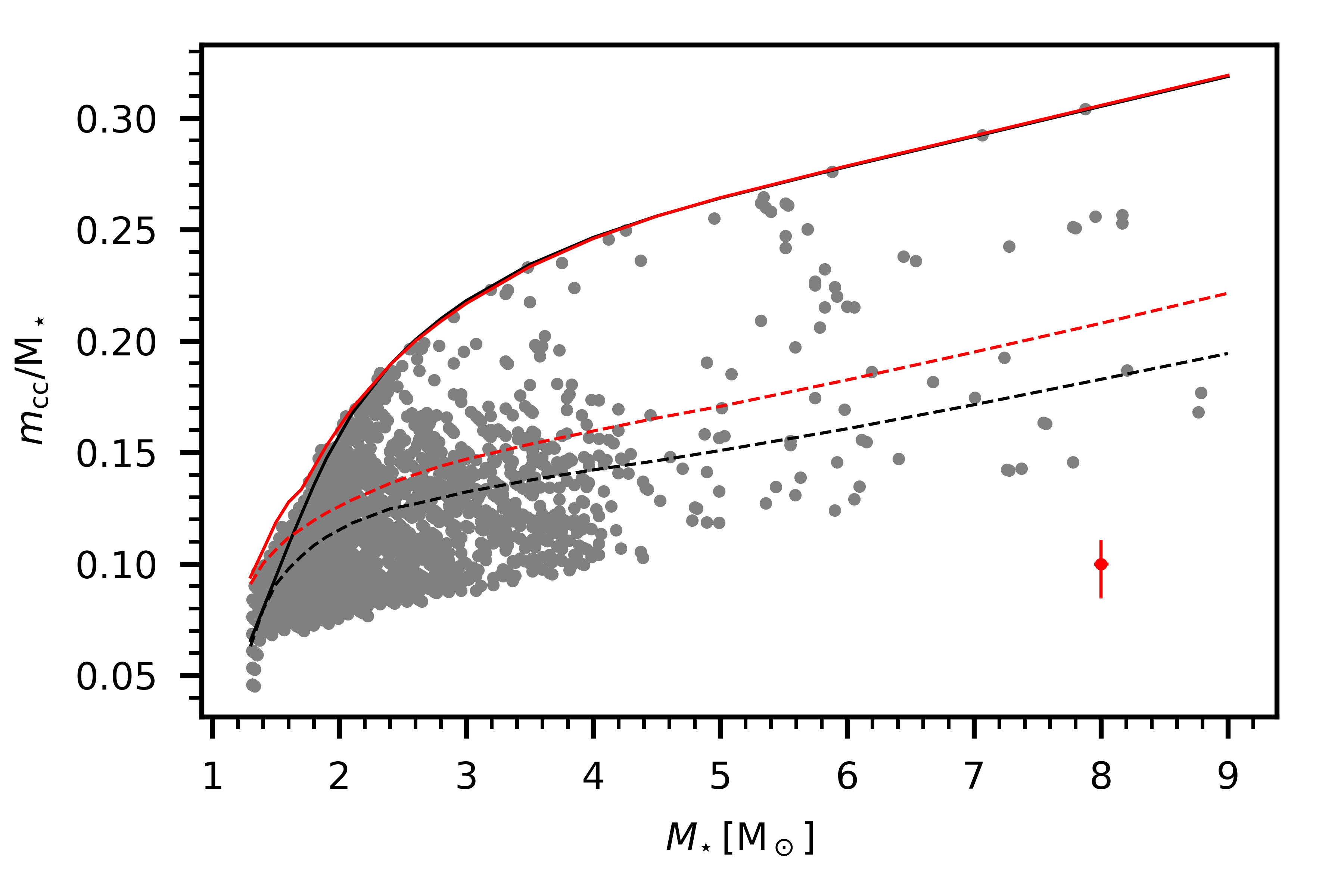}
    \caption{Relation between the inferred stellar mass and fractional
      convective core mass of the stars in the sample. The solid black (red)
     line indicates the maximum core mass possible during the
      main sequence for $f_{\rm CBM} = 0.005$ (0.025). The dashed
      lines show the maximum core mass at $X_{\rm c}/X_{\rm ini} =
      0.4$ (same colour coding). Average uncertainties are indicated
      by the red data point.}
    \label{fig:mass_vs_mcc}
\end{figure}

\begin{figure}
    \centering
    \includegraphics[width = \columnwidth]{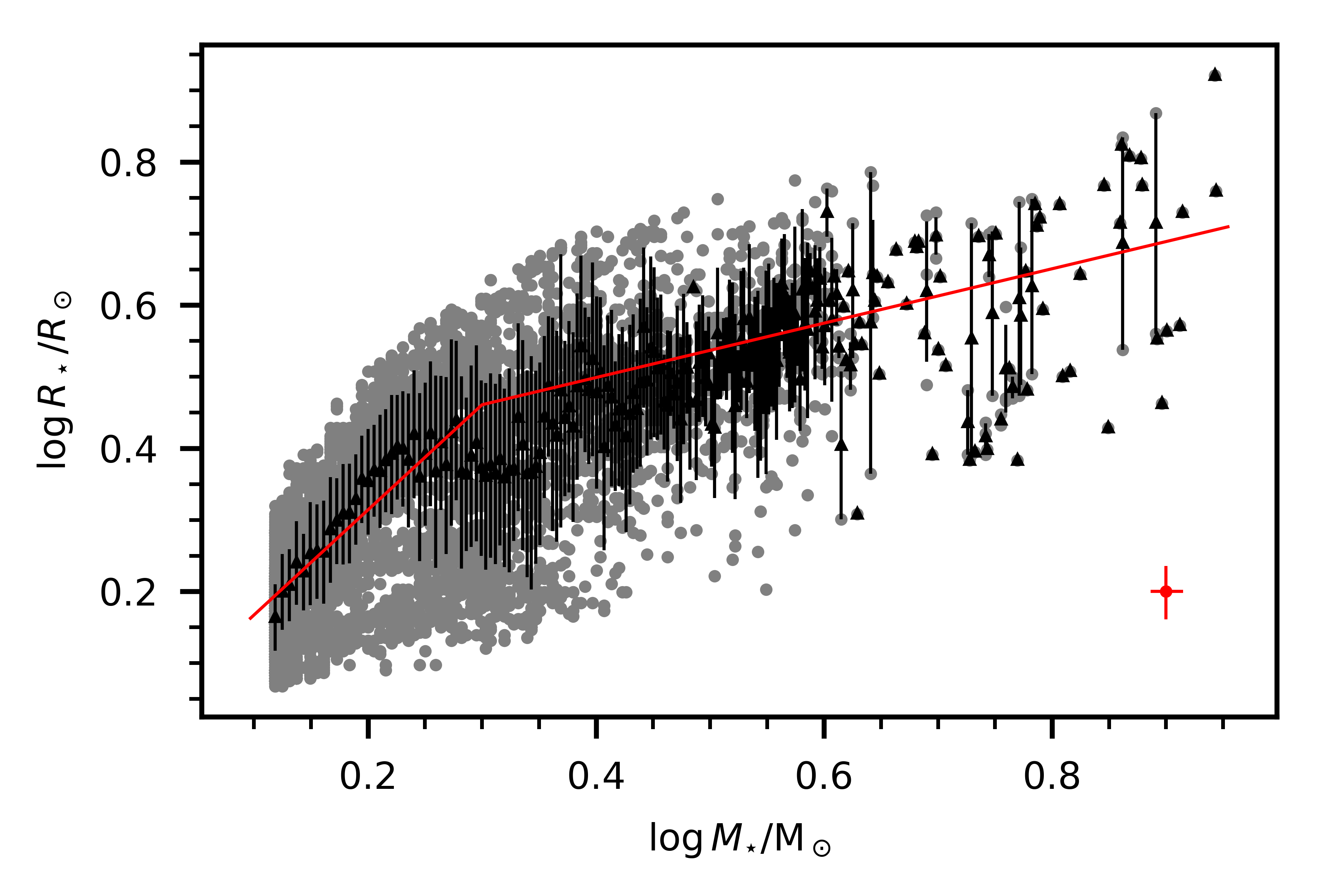}
    \caption{Relation between the inferred stellar mass and the
      inferred stellar radius (from models) of the stars in the
      sample. \edit{The solid red line indicates a two-component
        linear fit to the data (see text). The black triangles
        indicate bin averaged-values.} Average uncertainties are
      indicated by the red data point.}
    \label{fig:mass_vs_radius}
\end{figure}

\begin{figure}
    \centering
    \includegraphics[width = \columnwidth]{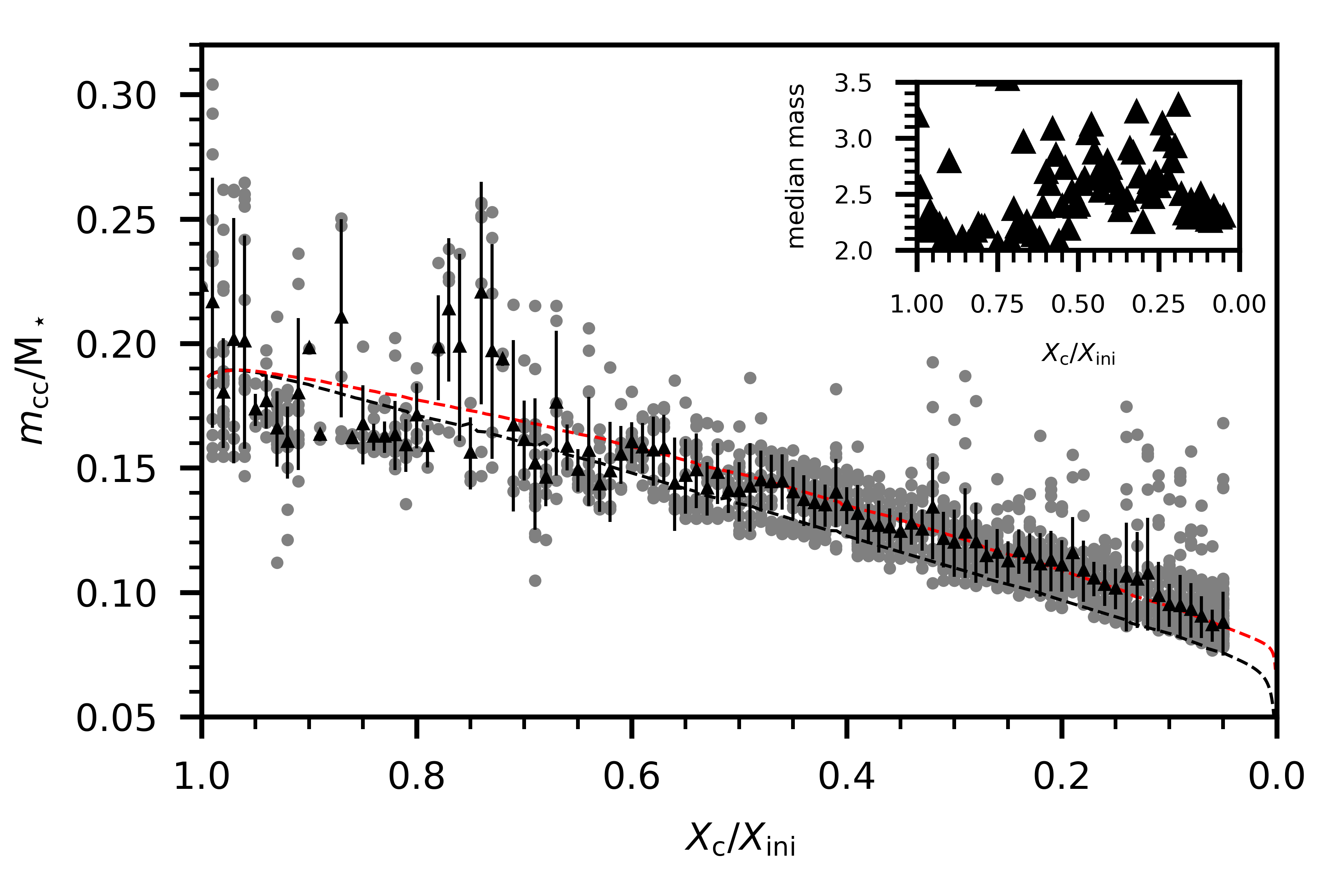}
    \caption{Relation between the inferred fraction \XcX and the
      convective core mass for stars $M_\star > 2{\rm M_\odot}$. The
      black triangles indicate the average per bin and standard
      deviation. Two models are shown for 2.4\Msun and $\omega_0 =
      0.05$, where the mass is representative of the median stellar
      mass per bin as shown in the inset. The black colour corresponds
      to $f_{\rm CBM} = 0.005$, the red colour to $f_{\rm CBM} =
      0.025$}
    \label{fig:xcx_vs_mcc}
\end{figure}

\begin{figure*}
    \centering
    \includegraphics[width = 0.97\textwidth]{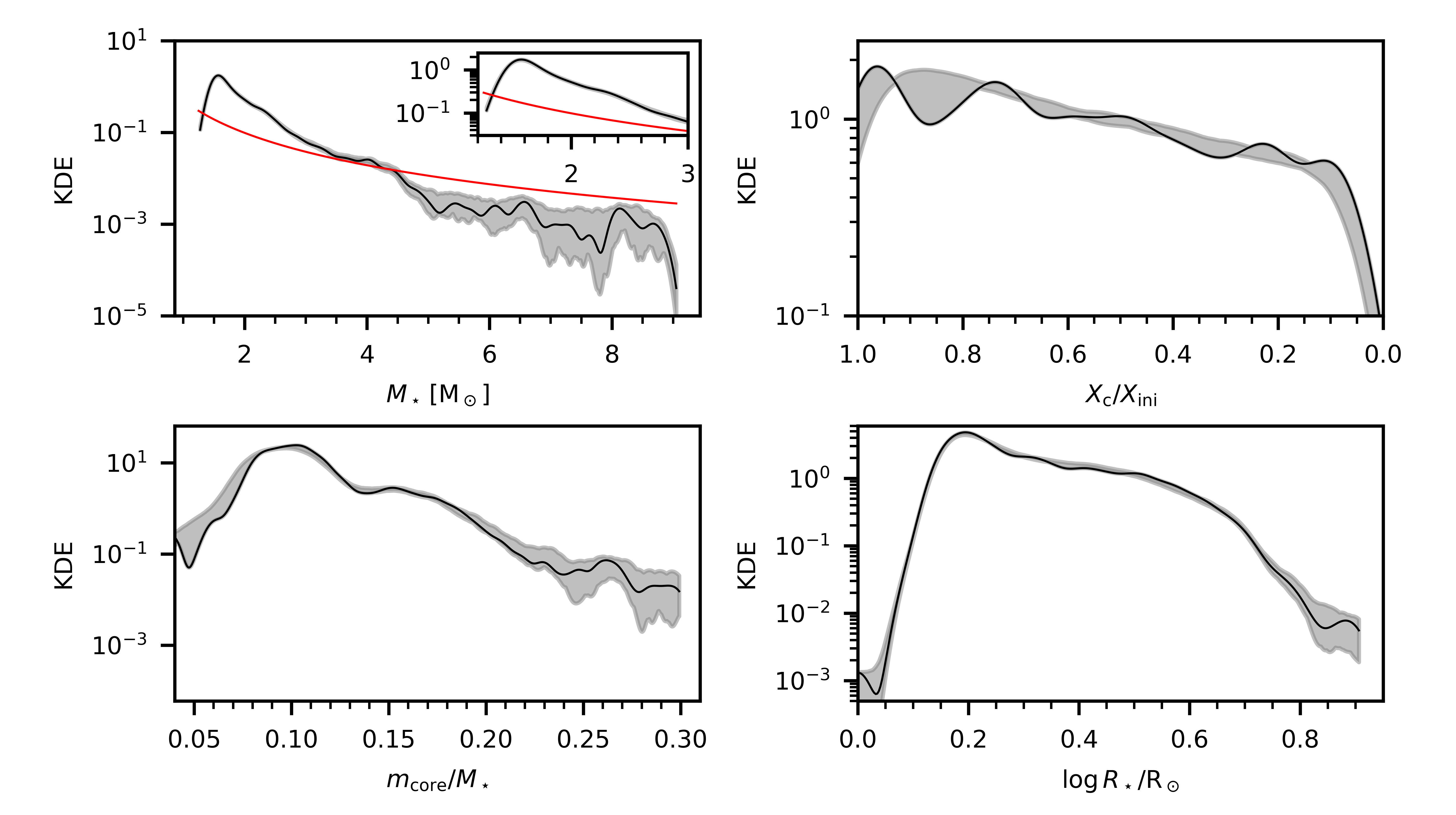}
    \caption{Same as Fig.~\ref{fig:distr_error}, but for $Z = 0.014$. }
    \label{fig:distr_error_Zsun}
\end{figure*}

The distribution of masses within the \gdor range \edit{has a similar
  shape compared to} the mass distribution derived by
\cite{Mombarg2024} for the 539 genuine \gdor stars observed by the
     {\it Kepler} mission. We do, however, find a peak in the number
     of stars around 1.4\Msun (top-left panel in
     Fig.~\ref{fig:distr_error}), which is slightly lower in mass than
     the peak around $\sim$1.5\Msun observed for the {\it Kepler\/}
     prograde dipole g-mode pulsators considered by
     \cite{Mombarg2024}. Furthermore, we observe a clear relation
     between the inferred mass and the observed luminosity, \edit{with
       $\log({L_\star}/{L_\odot}) = {4.0623(6)} \log\left
       ({M_\star}/{{\rm M}_\odot} \right) + 0.3468(2)$ as shown in
       Fig.~\ref{fig:mass_vs_lum}. This is a somewhat steeper
       mass--luminosity} relation than the one found by
     \citet{Yakut2007} based on eclipsing binaries with a B-type star
     component, namely $\log({L_\star}/{L_\odot}) = {3.724} \log\left
     ({M_\star}/{M_\odot} \right) + 0.162$. This is understood in
     terms of the increased importance of the radiation pressure with
     respect to the ideal gas law in the equation of state when going
     from F-type stars to B-type stars.

We find the majority of stars to form a plateau around 40 per cent of
their initial hydrogen mass fraction left in the convective core, with
a shape of a broad normal distribution around this peak value. The
fact that most stars appear to be mid-main sequence is dominated by
the \gdor stars in the parameter region where convective flux blocking is
active.

The bottom-left panel of Fig.~\ref{fig:distr_error} shows the
distributions of the fractional convective core masses ($m_{\rm
  cc}/M_\star$). The peak originates from the fact that the sample
contains both stars having an initially growing convective core and
stars with purely receding cores during the main sequence. We show the
correlation between the mass of the convective core and total stellar
mass in Fig.~\ref{fig:mass_vs_mcc}.  The solid black and red lines
show the maximum convective core mass as a function of total mass
occurring in the grid, for CMB with $f_{\rm CBM} = 0.005$ and $f_{\rm
  CBM} = 0.025$, respectively. The dashed lines show the location for
about half the main sequence, that is, at $X_{\rm c}/X_{\rm ini} =
0.4$.  We see that numerous stars occur between the black and red
lines, which reveals that a fraction of \gdor stars with a mass below
$\sim\,$1.8\Msun experience some level of CBM. However, most stars
have a convective core mass below the maximum for modest CBM ($f_{\rm
  CBM} = 0.005$), as was also found by \cite{Mombarg2024} for the
genuine {\it Kepler\/} \gdor sample. The maximum convective core mass
of stars above roughly 2\Msun is not dependent on the CBM efficiency
for the range of values we considered in our grids, but we recall that
we excluded envelope mixing in our model grids while this may be an
important ingredient to deduce the convective core mass of SPB stars
\citep{Pedersen2021,Pedersen2022a}. \edit{Asteroseismic modelling of
  oscillation frequencies is
  required to evaluate the cumulative effect of envelope mixing.}

The bottom-right of Fig.~\ref{fig:distr_error} shows the distributions
of the inferred stellar radii. Figure~\ref{fig:mass_vs_radius} shows
the mass against the radius. As is common, we find more scatter in the
mass--radius relation than in the mass--luminosity relation.  \edit{We
  perform a fit with two linear components, where we place the knee at
  $\log M_\star/{\rm M}_\odot = 0.3$ (2\Msun), as stars below this
  mass are less inflated because they have a thin outer convective
  layer, while those with a higher mass have a fully radiative
  envelope. This yields $R_\star \propto M_\star^{1.469(3)}$ on the
  lower-mass end, and $R_\star \propto M_\star^{0.380(4)}$ on the
  higher-mass end, where the knee is located at $\log R_\star/R_\odot
  = 0.4609(5)$.}

In Fig.~\ref{fig:xcx_vs_mcc} (and Fig.~\ref{fig:xcx_vs_mcc_Z140} for $Z=0.014$), we show the derived correlation between
the \edit{evolutionary stage} (via \XcX) and the convective-core mass. In this figure, we
only show stars above 2\Msun, as the \gdor stars are not equally
sampled in mass for each bin in \XcX, since the more massive ones tend
to be closer to the TAMS, while the less massive ones are closer to
the ZAMS. The downward trend of the core mass over time observed in
Fig.~\ref{fig:xcx_vs_mcc} is consistent with a constant value of \fov
along the main sequence, as can be seen from the models shown in this
figure. However, variations in the value of \fov along the main
sequence between the most extreme values used in our grids cannot be
ruled out. While we do appreciate that the trend shown in
Fig.~\ref{fig:xcx_vs_mcc} is to some extent enforced by the physics of
the evolution models used, we note that this physics is based on
\edit{detailed asteroseismic modelling studies} \citep[e.g.][]{Pedersen2021, Michielsen2021, Burssens2023, Mombarg2024}.

We repeat the exercise with a CNF trained on the grid with $Z =
0.014$. The results for the stars with $P_{\rm pred} \ge 0.5$ are
shown in Fig.~\ref{fig:distr_error_Zsun}. Our conclusions are
qualitatively the same for this grid modelling keeping in mind a tight
mass-metallicity relation, except that we now see an excess of stars
that fall below the ZAMS line. \edit{Using this assumption of the
  metallicity, we obtain a mass-luminosity relation
  $\log({L_\star}/{L_\odot}) = {4.1825(5)} \log\left ({M_\star}/{{\rm
      M}_\odot} \right) + 0.1057(2)$. For the mass-radius relation we
  find $R_\star \propto M_\star^{1.740(3)}$ on the lower-mass end, and
  $R_\star \propto M_\star^{0.623(3)}$ on the higher-mass end, where
  the knee is located at $\log R/R_\odot = 0.3945(3)$.} The
corresponding figures are shown in Appendix~\ref{ap:mass-lum}. We
expect most stars to have a true metallicity between the two extremes
considered here. The more metal-poor models give better consistency
for the observed pulsators as a population \edit{above the ZAMS}
based on the {\it Gaia\/}
effective temperatures and luminosities, while their membership to the
thin disk of the Milky Way would advocate for a more solar composition
\citep[see][for a detailed discussion]{deLaverny2024}. \edit{We
  quantify the effect of assuming an incorrect metallicity in
  Appendix~\ref{ap:method}.}

\section{Conclusions} \label{sect:conclusions}
In this paper, we have made use of conditional normalising flows
trained on two grids of rotating one-dimensional stellar evolution
models calibrated by asteroseismology of \gdor stars. Using this
method, we derived masses, \edit{evolutionary stages (on the main sequence)}, convective core masses, and radii of
\nsample new galactic g-mode pulsators. These form an all-sky sample
and were discovered from sparse {\it Gaia\/} DR3 light curves. All
stars were confirmed to be multiperiodic pulsators from TESS nominal
high-cadence space photometry \citep{HeyAerts2024}. This sample of
TESS-confirmed g-mode main-sequence pulsators offers an increase with
more than a factor 23 compared to the {\it Kepler\/} sample of
asteroseismically modelled main-sequence g-mode pulsators.

The parameter determination we provide here is based on the stars'
effective temperature and luminosity only, and hence represents
evolutionary parameters. These estimates cannot be compared with
\edit{parameters deduced from}
detailed asteroseismic modelling, but they offer initial estimates from the
homogeneously treated Gaia DR3 data. The two observables for each star
were coupled to a grid of stellar evolution models constructed with
input physics deduced from the sample of 611 asteroseismically
calibrated {\it Kepler\/} \gdor pulsators covering rotation rates from
almost zero to almost the critical rate.  The homogeneously determined
(convective core) masses and radii offer a valuable starting point for
future ensemble asteroseismic modelling of these g-mode pulsators,
once a sufficient number of their modes can be identified from
high-cadence TESS or future PLATO \citep{Rauer2024} space photometry.

\edit{Biases in the effective temperature of GSP-Phot used in this
  work might be present. The study by \cite{Avdeeva2024} report
  systematically higher effective temperatures measured by GSP-Phot
  compared those of the APOGEE survey. The largest discrepancies
  between GSP-Phot and APOGEE are, however, reported for $BP-RP$
  values between 2 and 4.5\,mag, whereas the stars in our sample have
  $BP-RP < 1.5$. A potential systematic overestimation of the
  effective temperature by GSP-Phot would lead to overestimated
  stellar masses and underestimated values for \XcX. On the other
  hand, \cite{Avdeeva2024} show that the effective temperatures
  from GSP-Phot and the GALAH survey are globally in good agreement
  (limited to effective temperatures below 8000\,K). Asteroseismic
  effective temperatures will be inferred from follow-up studies
  of this sample and may shed light on the systematics. }

The distributions of stellar masses of the sample of g-mode pulsators
are in general agreement with a Salpeter IMF, except that we find an
excess of stars within the mass range of the \gdor pulsators. This
suggests that flux blocking at the bottom of the thin convective
envelope is a highly effective excitation mechanism for g-mode
oscillations \citep{Guzik2000,Dupret2005}. On the other hand, g-mode
excitation by the $\kappa$-mechanism seems to occur effectively in stars
covering masses from about 2\Msun to about 9\Msun, \edit{but our
  sample is not representative for the stars with a mass above
  4\,M$_\odot$. As long as we do not have the full lists of detected
  identified oscillation modes per star, we cannot make more detailed
  inferences on the effectivity of the excitation mechanisms in terms
  of particular modes. Such inferences will be the subject of further
  studies.}

\begin{acknowledgements}
\edit{We thank the anonymous referee for the suggestions that have
  improved the presentation of our results. } JSGM acknowledges
funding the French Agence Nationale de la Recherche (ANR), under grant
MASSIF (ANR-21-CE31-0018-02).  The research leading to these results
has received funding from the KU\,Leuven Research Council (grant
C16/18/005: PARADISE) and from the European Research Council (ERC)
under the Horizon Europe programme (Synergy Grant agreement
N$^\circ$101071505: 4D-STAR).  While partially funded by the European
Union, views and opinions expressed are however those of the authors
only and do not necessarily reflect those of the European Union or the
European Research Council. Neither the European Union nor the granting
authority can be held responsible for them. This research has made use
of the \texttt{Numpy} \citep{Numpy}, \texttt{Scipy} \citep{2020SciPy}
and \texttt{Matplotlib} \citep{Matplotlib} software packages.
\end{acknowledgements}

\bibliographystyle{aa}
\bibliography{RMRA.bib}

\begin{thebibliography}{103}
\expandafter\ifx\csname natexlab\endcsname\relax\def\natexlab#1{#1}\fi

\bibitem[{{Aerts}(2021)}]{Aerts2021}
{Aerts}, C. 2021, Reviews of Modern Physics, 93, 015001

\bibitem[{{Aerts} {et~al.}(2010){Aerts}, {Christensen-Dalsgaard}, \& {Kurtz}}]{Aerts2010}
{Aerts}, C., {Christensen-Dalsgaard}, J., \& {Kurtz}, D.~W. 2010, {Asteroseismology, Springer-Verlag Heidelberg}

\bibitem[{{Aerts} {et~al.}(1999){Aerts}, {De Cat}, {Peeters}, {Decin}, {De Ridder}, {Kolenberg}, {Meeus}, {Van Winckel}, {Cuypers}, \& {Waelkens}}]{Aerts1999}
{Aerts}, C., {De Cat}, P., {Peeters}, E., {et~al.} 1999, \aap, 343, 872

\bibitem[{{Aerts} {et~al.}(2019){Aerts}, {Mathis}, \& {Rogers}}]{Aerts2019-araa}
{Aerts}, C., {Mathis}, S., \& {Rogers}, T.~M. 2019, \araa, 57, 35

\bibitem[{{Aerts} {et~al.}(2023){Aerts}, {Molenberghs}, \& {De Ridder}}]{Aerts2023}
{Aerts}, C., {Molenberghs}, G., \& {De Ridder}, J. 2023, \aap, 672, A183

\bibitem[{{Aerts} {et~al.}(2018){Aerts}, {Molenberghs}, {Michielsen}, {Pedersen}, {Bj{\"o}rklund}, {Johnston}, {Mombarg}, {Bowman}, {Buysschaert}, {P{\'a}pics}, {Sekaran}, {Sundqvist}, {Tkachenko}, {Truyaert}, {Van Reeth}, \& {Vermeyen}}]{Aerts2018}
{Aerts}, C., {Molenberghs}, G., {Michielsen}, M., {et~al.} 2018, \apjs, 237, 15

\bibitem[{{Asplund} {et~al.}(2009){Asplund}, {Grevesse}, {Sauval}, \& {Scott}}]{Asplund2009}
{Asplund}, M., {Grevesse}, N., {Sauval}, A.~J., \& {Scott}, P. 2009, \araa, 47, 481

\bibitem[{{Avdeeva} {et~al.}(2024){Avdeeva}, {Kovaleva}, {Malkov}, \& {Zhao}}]{Avdeeva2024}
{Avdeeva}, A.~S., {Kovaleva}, D.~A., {Malkov}, O.~Y., \& {Zhao}, G. 2024, \mnras, 527, 7382

\bibitem[{{Balona}(2024)}]{Balona2024}
{Balona}, L.~A. 2024, \apj, submitted, arXiv:2310.09805

\bibitem[{{Balona} \& {Ozuyar}(2020)}]{Balona2020}
{Balona}, L.~A. \& {Ozuyar}, D. 2020, \mnras, 493, 5871

\bibitem[{{Blomme} {et~al.}(2010){Blomme}, {Debosscher}, {De Ridder}, {Aerts}, {Gilliland}, {Christensen-Dalsgaard}, {Kjeldsen}, {Brown}, {Borucki}, {Koch}, {Jenkins}, {Kurtz}, {Stello}, {Stevens}, {Suran}, \& {Derekas}}]{Blomme2010}
{Blomme}, J., {Debosscher}, J., {De Ridder}, J., {et~al.} 2010, \apjl, 713, L204

\bibitem[{{Bouabid} {et~al.}(2013){Bouabid}, {Dupret}, {Salmon}, {Montalb{\'a}n}, {Miglio}, \& {Noels}}]{Bouabid2013}
{Bouabid}, M.~P., {Dupret}, M.~A., {Salmon}, S., {et~al.} 2013, \mnras, 429, 2500

\bibitem[{{Bowman}(2017)}]{Bowman_BOOK}
{Bowman}, D.~M. 2017, {Amplitude Modulation of Pulsation Modes in Delta Scuti Stars} (Springer International Publishing)

\bibitem[{{Burssens} {et~al.}(2023){Burssens}, {Bowman}, {Michielsen}, {Sim{\'o}n-D{\'\i}az}, {Aerts}, {Vanlaer}, {Banyard}, {Nardetto}, {Townsend}, {Handler}, {Mombarg}, {Vanderspek}, \& {Ricker}}]{Burssens2023}
{Burssens}, S., {Bowman}, D.~M., {Michielsen}, M., {et~al.} 2023, Nature Astronomy, 7, 913

\bibitem[{{Chaboyer} \& {Zahn}(1992)}]{Chaboyer1992}
{Chaboyer}, B. \& {Zahn}, J.~P. 1992, \aap, 253, 173

\bibitem[{{Choi} {et~al.}(2016){Choi}, {Dotter}, {Conroy}, {Cantiello}, {Paxton}, \& {Johnson}}]{Choi2016}
{Choi}, J., {Dotter}, A., {Conroy}, C., {et~al.} 2016, \apj, 823, 102

\bibitem[{{Cox} \& {Giuli}(1968)}]{Cox1968}
{Cox}, J.~P. \& {Giuli}, R.~T. 1968, {Principles of stellar structure}

\bibitem[{{De Cat} \& {Aerts}(2002)}]{DeCatAerts2002}
{De Cat}, P. \& {Aerts}, C. 2002, \aap, 393, 965

\bibitem[{{de Laverny} {et~al.}(2024){de Laverny}, {Recio-Blanco}, {Aerts}, \& {Palicio}}]{deLaverny2024}
{de Laverny}, P., {Recio-Blanco}, A., {Aerts}, C., \& {Palicio}, P.~A. 2024, \aap, in press, arXiv:2409.03361

\bibitem[{{Debosscher} {et~al.}(2011){Debosscher}, {Blomme}, {Aerts}, \& {De Ridder}}]{Debosscher2011}
{Debosscher}, J., {Blomme}, J., {Aerts}, C., \& {De Ridder}, J. 2011, \aap, 529, A89

\bibitem[{{Debosscher} {et~al.}(2009){Debosscher}, {Sarro}, {L{\'o}pez}, {Deleuil}, {Aerts}, {Auvergne}, {Baglin}, {Baudin}, {Chadid}, {Charpinet}, {Cuypers}, {De Ridder}, {Garrido}, {Hubert}, {Janot-Pacheco}, {Jorda}, {Kaiser}, {Kallinger}, {Kollath}, {Maceroni}, {Mathias}, {Michel}, {Moutou}, {Neiner}, {Ollivier}, {Samadi}, {Solano}, {Surace}, {Vandenbussche}, \& {Weiss}}]{Debosscher2009}
{Debosscher}, J., {Sarro}, L.~M., {L{\'o}pez}, M., {et~al.} 2009, \aap, 506, 519

\bibitem[{{Degroote} {et~al.}(2010){Degroote}, {Aerts}, {Baglin}, {Miglio}, {Briquet}, {Noels}, {Niemczura}, {Montalban}, {Bloemen}, {Oreiro}, {Vu{\v{c}}kovi{\'c}}, {Smolders}, {Auvergne}, {Baudin}, {Catala}, \& {Michel}}]{Degroote2010}
{Degroote}, P., {Aerts}, C., {Baglin}, A., {et~al.} 2010, \nat, 464, 259

\bibitem[{{Degroote} {et~al.}(2009){Degroote}, {Aerts}, {Ollivier}, {Miglio}, {Debosscher}, {Cuypers}, {Briquet}, {Montalb{\'a}n}, {Thoul}, {Noels}, {De Cat}, {Balaguer-N{\'u}{\~n}ez}, {Maceroni}, {Ribas}, {Auvergne}, {Baglin}, {Deleuil}, {Weiss}, {Jorda}, {Baudin}, \& {Samadi}}]{Degroote2009a}
{Degroote}, P., {Aerts}, C., {Ollivier}, M., {et~al.} 2009, \aap, 506, 471

\bibitem[{{Dupret} {et~al.}(2005){Dupret}, {Grigahc{\`e}ne}, {Garrido}, {Gabriel}, \& {Scuflaire}}]{Dupret2005}
{Dupret}, M.~A., {Grigahc{\`e}ne}, A., {Garrido}, R., {Gabriel}, M., \& {Scuflaire}, R. 2005, \aap, 435, 927

\bibitem[{{Eyer} {et~al.}(2023){Eyer}, {Audard}, {Holl}, {Rimoldini}, {Carnerero}, {Clementini}, {De Ridder}, {Distefano}, {Evans}, {Gavras}, {Gomel}, {Lebzelter}, {Marton}, {Mowlavi}, {Panahi}, {Ripepi}, {Wyrzykowski}, {Nienartowicz}, {Jevardat de Fombelle}, {Lecoeur-Taibi}, {Rohrbasser}, {Riello}, {Garc{\'\i}a-Lario}, {Lanzafame}, {Mazeh}, {Raiteri}, {Zucker}, {{\'A}brah{\'a}m}, {Aerts}, {Aguado}, {Anderson}, {Bashi}, {Binnenfeld}, {Faigler}, {Garofalo}, {Karbevska}, {K{\'o}sp{\'a}l}, {Kruszy{\'n}ska}, {Kun}, {Lanza}, {Leccia}, {Marconi}, {Messina}, {Molinaro}, {Moln{\'a}r}, {Muraveva}, {Musella}, {Nagy}, {Pagano}, {Palaversa}, {Plachy}, {Pr{\v{s}}a}, {Rybicki}, {Shahaf}, {Szabados}, {Szegedi-Elek}, {Trabucchi}, {Barblan}, {Grenon}, {Roelens}, \& {S{\"u}veges}}]{Eyer2023}
{Eyer}, L., {Audard}, M., {Holl}, B., {et~al.} 2023, \aap, 674, A13

\bibitem[{{Fritzewski} {et~al.}(2024{\natexlab{a}}){Fritzewski}, {Aerts}, {Mombarg}, {Gossage}, \& {Van Reeth}}]{Fritzewski2024b}
{Fritzewski}, D.~J., {Aerts}, C., {Mombarg}, J.~S.~G., {Gossage}, S., \& {Van Reeth}, T. 2024{\natexlab{a}}, \aap, 684, A112

\bibitem[{{Fritzewski} {et~al.}(2024{\natexlab{b}}){Fritzewski}, {Van Reeth}, {Aerts}, {Van Beeck}, {Gossage}, \& {Li}}]{Fritzewski2024a}
{Fritzewski}, D.~J., {Van Reeth}, T., {Aerts}, C., {et~al.} 2024{\natexlab{b}}, \aap, 681, A13

\bibitem[{{Fuller} {et~al.}(2019){Fuller}, {Piro}, \& {Jermyn}}]{Fuller2019}
{Fuller}, J., {Piro}, A.~L., \& {Jermyn}, A.~S. 2019, \mnras, 485, 3661

\bibitem[{{Gaia Collaboration} {et~al.}(2018){Gaia Collaboration}, {Brown}, {Vallenari}, {Prusti}, {de Bruijne}, {Babusiaux}, \& {Bailer-Jones}}]{Brown2018}
{Gaia Collaboration}, {Brown}, A.~G.~A., {Vallenari}, A., {et~al.} 2018, \aap, 616, A1

\bibitem[{{Gaia Collaboration} {et~al.}(2016{\natexlab{a}}){Gaia Collaboration}, {Brown}, {Vallenari}, {Prusti}, \& {de Bruijne}}]{Brown2016}
{Gaia Collaboration}, {Brown}, A.~G.~A., {Vallenari}, A., {Prusti}, T., \& {de Bruijne}, J.~H.~J. e.~a. 2016{\natexlab{a}}, \aap, 595, A2

\bibitem[{{Gaia Collaboration} {et~al.}(2023{\natexlab{a}}){Gaia Collaboration}, {De Ridder}, {Ripepi}, {Aerts}, {Palaversa}, \& {Eyer}}]{DeRidder2023}
{Gaia Collaboration}, {De Ridder}, J., {Ripepi}, V., {et~al.} 2023{\natexlab{a}}, \aap, 674, A36

\bibitem[{{Gaia Collaboration} {et~al.}(2016{\natexlab{b}}){Gaia Collaboration}, {Prusti}, {de Bruijne}, {Brown}, \& {Vallenari}}]{Prusti2016}
{Gaia Collaboration}, {Prusti}, T., {de Bruijne}, J.~H.~J., {Brown}, A.~G.~A., \& {Vallenari}, A. e.~a. 2016{\natexlab{b}}, \aap, 595, A1

\bibitem[{{Gaia Collaboration} {et~al.}(2023{\natexlab{b}}){Gaia Collaboration}, {Vallenari}, {Brown}, {Prusti}, \& {de Bruijne}}]{Vallenari2023}
{Gaia Collaboration}, {Vallenari}, A., {Brown}, A.~G.~A., {Prusti}, T., \& {de Bruijne}, J.~H.~J. e.~a. 2023{\natexlab{b}}, \aap, 674, A1

\bibitem[{{Garcia} {et~al.}(2022{\natexlab{a}}){Garcia}, {Van Reeth}, {De Ridder}, \& {Aerts}}]{Garcia2022b}
{Garcia}, S., {Van Reeth}, T., {De Ridder}, J., \& {Aerts}, C. 2022{\natexlab{a}}, \aap, 668, A137

\bibitem[{{Garcia} {et~al.}(2022{\natexlab{b}}){Garcia}, {Van Reeth}, {De Ridder}, {Tkachenko}, {IJspeert}, \& {Aerts}}]{Garcia2022a}
{Garcia}, S., {Van Reeth}, T., {De Ridder}, J., {et~al.} 2022{\natexlab{b}}, \aap, 662, A82

\bibitem[{{Gebruers} {et~al.}(2021){Gebruers}, {Straumit}, {Tkachenko}, {Mombarg}, {Pedersen}, {Van Reeth}, {Li}, {Lampens}, {Escorza}, {Bowman}, {De Cat}, {Vermeylen}, {Bodensteiner}, {Rix}, \& {Aerts}}]{Gebruers2021}
{Gebruers}, S., {Straumit}, I., {Tkachenko}, A., {et~al.} 2021, \aap, 650, A151

\bibitem[{{Gilliland} {et~al.}(2010){Gilliland}, {Brown}, {Christensen-Dalsgaard}, {Kjeldsen}, {Aerts}, {Appourchaux}, {Basu}, {Bedding}, {Chaplin}, {Cunha}, {De Cat}, {De Ridder}, {Guzik}, {Handler}, {Kawaler}, {Kiss}, {Kolenberg}, {Kurtz}, {Metcalfe}, {Monteiro}, {Szab{\'o}}, {Arentoft}, {Balona}, {Debosscher}, {Elsworth}, {Quirion}, {Stello}, {Su{\'a}rez}, {Borucki}, {Jenkins}, {Koch}, {Kondo}, {Latham}, {Rowe}, \& {Steffen}}]{Gilliland2010}
{Gilliland}, R.~L., {Brown}, T.~M., {Christensen-Dalsgaard}, J., {et~al.} 2010, \pasp, 122, 131

\bibitem[{{Guzik} {et~al.}(2000){Guzik}, {Kaye}, {Bradley}, {Cox}, \& {Neuforge}}]{Guzik2000}
{Guzik}, J.~A., {Kaye}, A.~B., {Bradley}, P.~A., {Cox}, A.~N., \& {Neuforge}, C. 2000, \apjl, 542, L57

\bibitem[{Han \& Brandt(2023)}]{HanBrandt2023}
Han, T. \& Brandt, T.~D. 2023, The Astronomical Journal, 165, 71

\bibitem[{{Handler}(1999)}]{Handler1999}
{Handler}, G. 1999, \mnras, 309, L19

\bibitem[{{Hey} \& {Aerts}(2024)}]{HeyAerts2024}
{Hey}, D. \& {Aerts}, C. 2024, \aap, 688, A93

\bibitem[{{Hon} {et~al.}(2024){Hon}, {Li}, \& {Ong}}]{Hon2024}
{Hon}, M., {Li}, Y., \& {Ong}, J. 2024, \apj, 973, 154

\bibitem[{Hunter(2007)}]{Matplotlib}
Hunter, J.~D. 2007, Computing in Science Engineering, 9, 90

\bibitem[{{Jermyn} {et~al.}(2023){Jermyn}, {Bauer}, {Schwab}, {Farmer}, {Ball}, {Bellinger}, {Dotter}, {Joyce}, {Marchant}, {Mombarg}, {Wolf}, {Sunny Wong}, {Cinquegrana}, {Farrell}, {Smolec}, {Thoul}, {Cantiello}, {Herwig}, {Toloza}, {Bildsten}, {Townsend}, \& {Timmes}}]{Jermyn2023}
{Jermyn}, A.~S., {Bauer}, E.~B., {Schwab}, J., {et~al.} 2023, \apjs, 265, 15

\bibitem[{{Kaye} {et~al.}(1999){Kaye}, {Handler}, {Krisciunas}, {Poretti}, \& {Zerbi}}]{Kaye1999}
{Kaye}, A.~B., {Handler}, G., {Krisciunas}, K., {Poretti}, E., \& {Zerbi}, F.~M. 1999, \pasp, 111, 840

\bibitem[{{Kurtz}(2022)}]{Kurtz2022}
{Kurtz}, D.~W. 2022, \araa, 60, 31

\bibitem[{{Li} {et~al.}(2024){Li}, {Aerts}, {Bedding}, {Fritzewski}, {Murphy}, {Van Reeth}, {Montet}, {Jian}, {Mombarg}, {Gossage}, \& {Sreenivas}}]{GangLi2024}
{Li}, G., {Aerts}, C., {Bedding}, T.~R., {et~al.} 2024, \aap, 686, A142

\bibitem[{{Li} {et~al.}(2020){Li}, {Van Reeth}, {Bedding}, {Murphy}, {Antoci}, {Ouazzani}, \& {Barbara}}]{GangLi2020}
{Li}, G., {Van Reeth}, T., {Bedding}, T.~R., {et~al.} 2020, \mnras, 491, 3586

\bibitem[{{Michielsen} {et~al.}(2021){Michielsen}, {Aerts}, \& {Bowman}}]{Michielsen2021}
{Michielsen}, M., {Aerts}, C., \& {Bowman}, D.~M. 2021, \aap, 650, A175

\bibitem[{{Miglio} {et~al.}(2008){Miglio}, {Montalb{\'a}n}, {Noels}, \& {Eggenberger}}]{Miglio2008}
{Miglio}, A., {Montalb{\'a}n}, J., {Noels}, A., \& {Eggenberger}, P. 2008, \mnras, 386, 1487

\bibitem[{{Mombarg}(2023)}]{Mombarg2023}
{Mombarg}, J.~S.~G. 2023, \aap, 677, A63

\bibitem[{{Mombarg} {et~al.}(2024){Mombarg}, {Aerts}, \& {Molenberghs}}]{Mombarg2024}
{Mombarg}, J.~S.~G., {Aerts}, C., \& {Molenberghs}, G. 2024, \aap, 685, A21

\bibitem[{{Mombarg} {et~al.}(2021){Mombarg}, {Van Reeth}, \& {Aerts}}]{Mombarg2021}
{Mombarg}, J.~S.~G., {Van Reeth}, T., \& {Aerts}, C. 2021, \aap, 650, A58

\bibitem[{{Mombarg} {et~al.}(2019){Mombarg}, {Van Reeth}, {Pedersen}, {Molenberghs}, {Bowman}, {Johnston}, {Tkachenko}, \& {Aerts}}]{Mombarg2019}
{Mombarg}, J.~S.~G., {Van Reeth}, T., {Pedersen}, M.~G., {et~al.} 2019, \mnras, 485, 3248

\bibitem[{{Moravveji} {et~al.}(2015){Moravveji}, {Aerts}, {P{\'a}pics}, {Triana}, \& {Vandoren}}]{Moravveji2015}
{Moravveji}, E., {Aerts}, C., {P{\'a}pics}, P.~I., {Triana}, S.~A., \& {Vandoren}, B. 2015, \aap, 580, A27

\bibitem[{{Moravveji} {et~al.}(2016){Moravveji}, {Townsend}, {Aerts}, \& {Mathis}}]{Moravveji2016}
{Moravveji}, E., {Townsend}, R.~H.~D., {Aerts}, C., \& {Mathis}, S. 2016, \apj, 823, 130

\bibitem[{{Mowlavi} {et~al.}(2013){Mowlavi}, {Barblan}, {Saesen}, \& {Eyer}}]{Mowlavi2013}
{Mowlavi}, N., {Barblan}, F., {Saesen}, S., \& {Eyer}, L. 2013, \aap, 554, A108

\bibitem[{{Mowlavi} {et~al.}(2016){Mowlavi}, {Saesen}, {Semaan}, {Eggenberger}, {Barblan}, {Eyer}, {Ekstr{\"o}m}, \& {Georgy}}]{Mowlavi2016}
{Mowlavi}, N., {Saesen}, S., {Semaan}, T., {et~al.} 2016, \aap, 595, L1

\bibitem[{{Moyano} {et~al.}(2024){Moyano}, {Eggenberger}, \& {Salmon}}]{Moyano2024}
{Moyano}, F.~D., {Eggenberger}, P., \& {Salmon}, S.~J.~A.~J. 2024, \aap, 681, L16

\bibitem[{{Nieva} \& {Przybilla}(2012)}]{Nieva2012}
{Nieva}, M.~F. \& {Przybilla}, N. 2012, \aap, 539, A143

\bibitem[{{Ouazzani} {et~al.}(2019){Ouazzani}, {Marques}, {Goupil}, {Christophe}, {Antoci}, {Salmon}, \& {Ballot}}]{Ouazzani2019}
{Ouazzani}, R.~M., {Marques}, J.~P., {Goupil}, M.~J., {et~al.} 2019, \aap, 626, A121

\bibitem[{{Ouazzani} {et~al.}(2017){Ouazzani}, {Salmon}, {Antoci}, {Bedding}, {Murphy}, \& {Roxburgh}}]{Ouazzani2017}
{Ouazzani}, R.-M., {Salmon}, S.~J.~A.~J., {Antoci}, V., {et~al.} 2017, \mnras, 465, 2294

\bibitem[{{P{\'a}pics} {et~al.}(2012){P{\'a}pics}, {Briquet}, {Baglin}, {Poretti}, {Aerts}, {Degroote}, {Tkachenko}, {Morel}, {Zima}, {Niemczura}, {Rainer}, {Hareter}, {Baudin}, {Catala}, {Michel}, {Samadi}, \& {Auvergne}}]{Papics2012}
{P{\'a}pics}, P.~I., {Briquet}, M., {Baglin}, A., {et~al.} 2012, \aap, 542, A55

\bibitem[{{P{\'a}pics} {et~al.}(2014){P{\'a}pics}, {Moravveji}, {Aerts}, {Tkachenko}, {Triana}, {Bloemen}, \& {Southworth}}]{Papics2014}
{P{\'a}pics}, P.~I., {Moravveji}, E., {Aerts}, C., {et~al.} 2014, \aap, 570, A8

\bibitem[{{P{\'a}pics} {et~al.}(2017){P{\'a}pics}, {Tkachenko}, {Van Reeth}, {Aerts}, {Moravveji}, {Van de Sande}, {De Smedt}, {Bloemen}, {Southworth}, {Debosscher}, {Niemczura}, \& {Gameiro}}]{Papics2017}
{P{\'a}pics}, P.~I., {Tkachenko}, A., {Van Reeth}, T., {et~al.} 2017, \aap, 598, A74

\bibitem[{{Paxton} {et~al.}(2011){Paxton}, {Bildsten}, {Dotter}, {Herwig}, {Lesaffre}, \& {Timmes}}]{Paxton2011}
{Paxton}, B., {Bildsten}, L., {Dotter}, A., {et~al.} 2011, \apjs, 192, 3

\bibitem[{{Paxton} {et~al.}(2013){Paxton}, {Cantiello}, {Arras}, {Bildsten}, {Brown}, {Dotter}, {Mankovich}, {Montgomery}, {Stello}, {Timmes}, \& {Townsend}}]{Paxton2013}
{Paxton}, B., {Cantiello}, M., {Arras}, P., {et~al.} 2013, \apjs, 208, 4

\bibitem[{{Paxton} {et~al.}(2015){Paxton}, {Marchant}, {Schwab}, {Bauer}, {Bildsten}, {Cantiello}, {Dessart}, {Farmer}, {Hu}, { }, {Townsend}, {Townsley}, \& {Timmes}}]{Paxton2015}
{Paxton}, B., {Marchant}, P., {Schwab}, J., {et~al.} 2015, \apjs, 220, 15

\bibitem[{{Paxton} {et~al.}(2018){Paxton}, {Schwab}, {Bauer}, {Bildsten}, {Blinnikov}, {Duffell}, {Farmer}, {Goldberg}, {Marchant}, {Sorokina}, {Thoul}, {Townsend}, \& {Timmes}}]{Paxton2018}
{Paxton}, B., {Schwab}, J., {Bauer}, E.~B., {et~al.} 2018, \apjs, 234, 34

\bibitem[{{Paxton} {et~al.}(2019){Paxton}, {Smolec}, {Schwab}, {Gautschy}, {Bildsten}, {Cantiello}, {Dotter}, {Farmer}, {Goldberg}, {Jermyn}, {Kanbur}, {Marchant}, {Thoul}, {Townsend}, {Wolf}, {Zhang}, \& {Timmes}}]{Paxton2019}
{Paxton}, B., {Smolec}, R., {Schwab}, J., {et~al.} 2019, \apjs, 243, 10

\bibitem[{{Pedersen}(2022{\natexlab{a}})}]{Pedersen2022b}
{Pedersen}, M.~G. 2022{\natexlab{a}}, \apj, 940, 49

\bibitem[{{Pedersen}(2022{\natexlab{b}})}]{Pedersen2022a}
{Pedersen}, M.~G. 2022{\natexlab{b}}, \apj, 930, 94

\bibitem[{{Pedersen} {et~al.}(2021){Pedersen}, {Aerts}, {P{\'a}pics}, {Michielsen}, {Gebruers}, {Rogers}, {Molenberghs}, {Burssens}, {Garcia}, \& {Bowman}}]{Pedersen2021}
{Pedersen}, M.~G., {Aerts}, C., {P{\'a}pics}, P.~I., {et~al.} 2021, Nature Astronomy, 5, 715

\bibitem[{{Rauer} {et~al.}(2024){Rauer}, {Aerts}, {Cabrera}, {Deleuil}, {Erikson}, {Gizon}, \& et~al.}]{Rauer2024}
{Rauer}, H., {Aerts}, C., {Cabrera}, J., {et~al.} 2024, Experimental Astronomy, submitted, arXiv:2406.05447

\bibitem[{{Rozet} {et~al.}(2023){Rozet}, {Divo}, \& S.}]{Zuko}
{Rozet}, F., {Divo}, F., \& S., S. 2023, {Zuko}: Normalizing flows in PyTorch

\bibitem[{{Saio} {et~al.}(2021){Saio}, {Takata}, {Lee}, {Li}, \& {Van Reeth}}]{Saio2021}
{Saio}, H., {Takata}, M., {Lee}, U., {Li}, G., \& {Van Reeth}, T. 2021, \mnras, 502, 5856

\bibitem[{{Salpeter}(1955)}]{Salpeter1955}
{Salpeter}, E.~E. 1955, \apj, 121, 161

\bibitem[{{Schmid} \& {Aerts}(2016)}]{Schmid2016}
{Schmid}, V.~S. \& {Aerts}, C. 2016, \aap, 592, A116

\bibitem[{{Seaton}(2005)}]{Seaton2005}
{Seaton}, M.~J. 2005, \mnras, 362, L1

\bibitem[{{Sekaran} {et~al.}(2021){Sekaran}, {Tkachenko}, {Johnston}, \& {Aerts}}]{Sekaran2021}
{Sekaran}, S., {Tkachenko}, A., {Johnston}, C., \& {Aerts}, C. 2021, \aap, 648, A91

\bibitem[{{Silverman}(1986)}]{Silverman1986}
{Silverman}, B.~W. 1986, {Density Estimation for Statistics and Data Analysis}, Vol.~26 (Monographs on Statistics and Applied Probability, Chapman and Hall, London)

\bibitem[{{Szewczuk} \& {Daszy{\'n}ska-Daszkiewicz}(2017)}]{Szewczuk2017}
{Szewczuk}, W. \& {Daszy{\'n}ska-Daszkiewicz}, J. 2017, \mnras, 469, 13

\bibitem[{{Szewczuk} \& {Daszy{\'n}ska-Daszkiewicz}(2018)}]{Szewczuk2018}
{Szewczuk}, W. \& {Daszy{\'n}ska-Daszkiewicz}, J. 2018, \mnras, 478, 2243

\bibitem[{{Szewczuk} {et~al.}(2021){Szewczuk}, {Walczak}, \& {Daszy{\'n}ska-Daszkiewicz}}]{Szewczuk2021}
{Szewczuk}, W., {Walczak}, P., \& {Daszy{\'n}ska-Daszkiewicz}, J. 2021, \mnras, 503, 5894

\bibitem[{{Szewczuk} {et~al.}(2022){Szewczuk}, {Walczak}, {Daszy{\'n}ska-Daszkiewicz}, \& {Mo{\'z}dzierski}}]{Szewczuk2022}
{Szewczuk}, W., {Walczak}, P., {Daszy{\'n}ska-Daszkiewicz}, J., \& {Mo{\'z}dzierski}, D. 2022, \mnras, 511, 1529

\bibitem[{{Tkachenko} {et~al.}(2013){Tkachenko}, {Aerts}, {Yakushechkin}, {Debosscher}, {Degroote}, {Bloemen}, {P{\'a}pics}, {de Vries}, {Lombaert}, {Hrudkova}, {Fr{\'e}mat}, {Raskin}, \& {Van Winckel}}]{Tkachenko2013}
{Tkachenko}, A., {Aerts}, C., {Yakushechkin}, A., {et~al.} 2013, \aap, 556, A52

\bibitem[{{Uytterhoeven} {et~al.}(2011){Uytterhoeven}, {Moya}, {Grigahc{\`e}ne}, {Guzik}, {Guti{\'e}rrez-Soto}, {Smalley}, {Handler}, {Balona}, {Niemczura}, {Fox Machado}, {Benatti}, {Chapellier}, {Tkachenko}, {Szab{\'o}}, {Su{\'a}rez}, {Ripepi}, {Pascual}, {Mathias}, {Mart{\'\i}n-Ru{\'\i}z}, {Lehmann}, {Jackiewicz}, {Hekker}, {Gruberbauer}, {Garc{\'\i}a}, {Dumusque}, {D{\'\i}az-Fraile}, {Bradley}, {Antoci}, {Roth}, {Leroy}, {Murphy}, {De Cat}, {Cuypers}, {Kjeldsen}, {Christensen-Dalsgaard}, {Breger}, {Pigulski}, {Kiss}, {Still}, {Thompson}, \& {van Cleve}}]{Uytterhoeven2011}
{Uytterhoeven}, K., {Moya}, A., {Grigahc{\`e}ne}, A., {et~al.} 2011, \aap, 534, A125

\bibitem[{{Van Beeck} {et~al.}(2021){Van Beeck}, {Bowman}, {Pedersen}, {Van Reeth}, {Van Hoolst}, \& {Aerts}}]{VanBeeck2021}
{Van Beeck}, J., {Bowman}, D.~M., {Pedersen}, M.~G., {et~al.} 2021, \aap, 655, A59

\bibitem[{{van der Walt} {et~al.}(2011){van der Walt}, {Colbert}, \& {Varoquaux}}]{Numpy}
{van der Walt}, S., {Colbert}, S.~C., \& {Varoquaux}, G. 2011, Computing in Science and Engineering, 13, 22

\bibitem[{{Van Reeth} {et~al.}(2018){Van Reeth}, {Mombarg}, {Mathis}, {Tkachenko}, {Fuller}, {Bowman}, {Buysschaert}, {Johnston}, {Garc{\'\i}a Hern{\'a}ndez}, {Goldstein}, {Townsend}, \& {Aerts}}]{VanReeth2018}
{Van Reeth}, T., {Mombarg}, J.~S.~G., {Mathis}, S., {et~al.} 2018, \aap, 618, A24

\bibitem[{{Van Reeth} {et~al.}(2016){Van Reeth}, {Tkachenko}, \& {Aerts}}]{VanReeth2016}
{Van Reeth}, T., {Tkachenko}, A., \& {Aerts}, C. 2016, \aap, 593, A120

\bibitem[{{Van Reeth} {et~al.}(2015{\natexlab{a}}){Van Reeth}, {Tkachenko}, {Aerts}, {P{\'a}pics}, {Degroote}, {Debosscher}, {Zwintz}, {Bloemen}, {De Smedt}, {Hrudkova}, {Raskin}, \& {Van Winckel}}]{VanReeth2015a}
{Van Reeth}, T., {Tkachenko}, A., {Aerts}, C., {et~al.} 2015{\natexlab{a}}, \aap, 574, A17

\bibitem[{{Van Reeth} {et~al.}(2015{\natexlab{b}}){Van Reeth}, {Tkachenko}, {Aerts}, {P{\'a}pics}, {Triana}, {Zwintz}, {Degroote}, {Debosscher}, {Bloemen}, {Schmid}, {De Smedt}, {Fremat}, {Fuentes}, {Homan}, {Hrudkova}, {Karjalainen}, {Lombaert}, {Nemeth}, {{\O}stensen}, {Van De Steene}, {Vos}, {Raskin}, \& {Van Winckel}}]{VanReeth2015b}
{Van Reeth}, T., {Tkachenko}, A., {Aerts}, C., {et~al.} 2015{\natexlab{b}}, \apjs, 218, 27

\bibitem[{{Verma} {et~al.}(2019){Verma}, {Raodeo}, {Basu}, {Silva Aguirre}, {Mazumdar}, {Mosumgaard}, {Lund}, \& {Ranadive}}]{Verma2019}
{Verma}, K., {Raodeo}, K., {Basu}, S., {et~al.} 2019, \mnras, 483, 4678

\bibitem[{Virtanen {et~al.}(2020)Virtanen, Gommers, Oliphant, Haberland, Reddy, Cournapeau, Burovski, Peterson, Weckesser, Bright, {van der Walt}, Brett, Wilson, Millman, Mayorov, Nelson, Jones, Kern, Larson, Carey, Polat, Feng, Moore, {VanderPlas}, Laxalde, Perktold, Cimrman, Henriksen, Quintero, Harris, Archibald, Ribeiro, Pedregosa, {van Mulbregt}, \& {SciPy 1.0 Contributors}}]{2020SciPy}
Virtanen, P., Gommers, R., Oliphant, T.~E., {et~al.} 2020, Nature Methods, 17, 261

\bibitem[{{Waelkens}(1991)}]{Waelkens1991}
{Waelkens}, C. 1991, \aap, 246, 453

\bibitem[{{Waelkens} {et~al.}(1998){Waelkens}, {Aerts}, {Kestens}, {Grenon}, \& {Eyer}}]{Waelkens1998}
{Waelkens}, C., {Aerts}, C., {Kestens}, E., {Grenon}, M., \& {Eyer}, L. 1998, \aap, 330, 215

\bibitem[{{Wu} \& {Li}(2019)}]{Wu2019}
{Wu}, T. \& {Li}, Y. 2019, \apj, 881, 86

\bibitem[{{Wu} {et~al.}(2018){Wu}, {Li}, \& {Deng}}]{Wu2018}
{Wu}, T., {Li}, Y., \& {Deng}, Z.-m. 2018, \apj, 867, 47

\bibitem[{{Xiong} {et~al.}(2016){Xiong}, {Deng}, {Zhang}, \& {Wang}}]{Xiong2016}
{Xiong}, D.~R., {Deng}, L., {Zhang}, C., \& {Wang}, K. 2016, \mnras, 457, 3163

\bibitem[{{Yakut} {et~al.}(2007){Yakut}, {Aerts}, \& {Morel}}]{Yakut2007}
{Yakut}, K., {Aerts}, C., \& {Morel}, T. 2007, \aap, 467, 647

\bibitem[{{Zahn}(1992)}]{Zahn1992}
{Zahn}, J.~P. 1992, \aap, 265, 115

\bibitem[{{Zwintz} \& {Steindl}(2022)}]{Zwintz2022}
{Zwintz}, K. \& {Steindl}, T. 2022, Frontiers in Astronomy and Space Sciences, 9, 914738

\end{thebibliography}

\begin{appendix}
\section{Results for solar metallicity grid} \label{ap:mass-lum}
In this appendix, we show the derived mass-luminosity relation, mass-radius relation, and \XcX-core-mass relation, when we assume a solar metallicity of $Z=0.014$.
\begin{figure}[htb]
    \centering
    \includegraphics[width = \columnwidth]{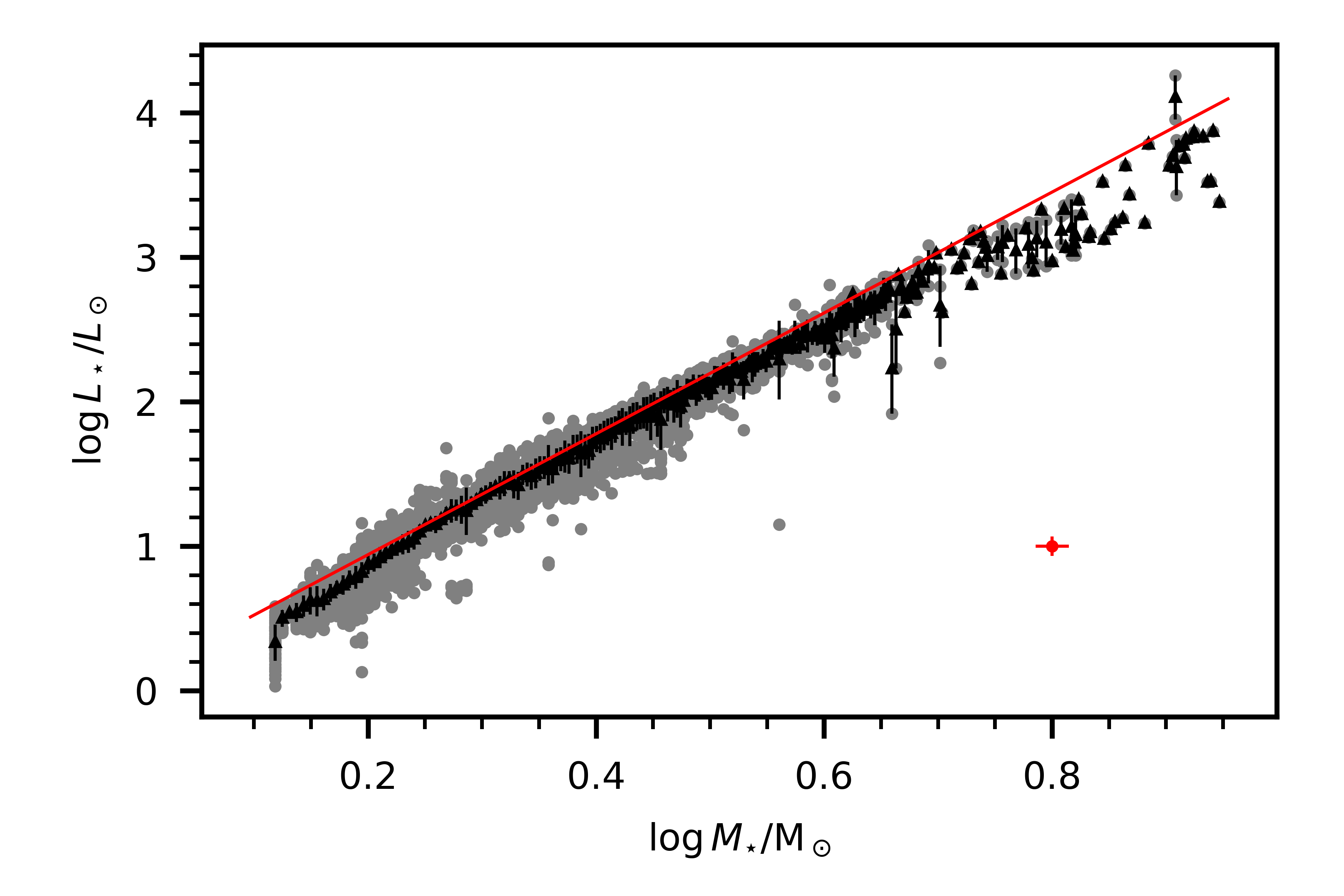}
    \caption{Same as Fig.~\ref{fig:mass_vs_lum}, but for $Z = 0.014$.}
    \label{fig:mass_vs_lum_Z140}
\end{figure}

\begin{figure}[htb]
    \centering
    \includegraphics[width = \columnwidth]{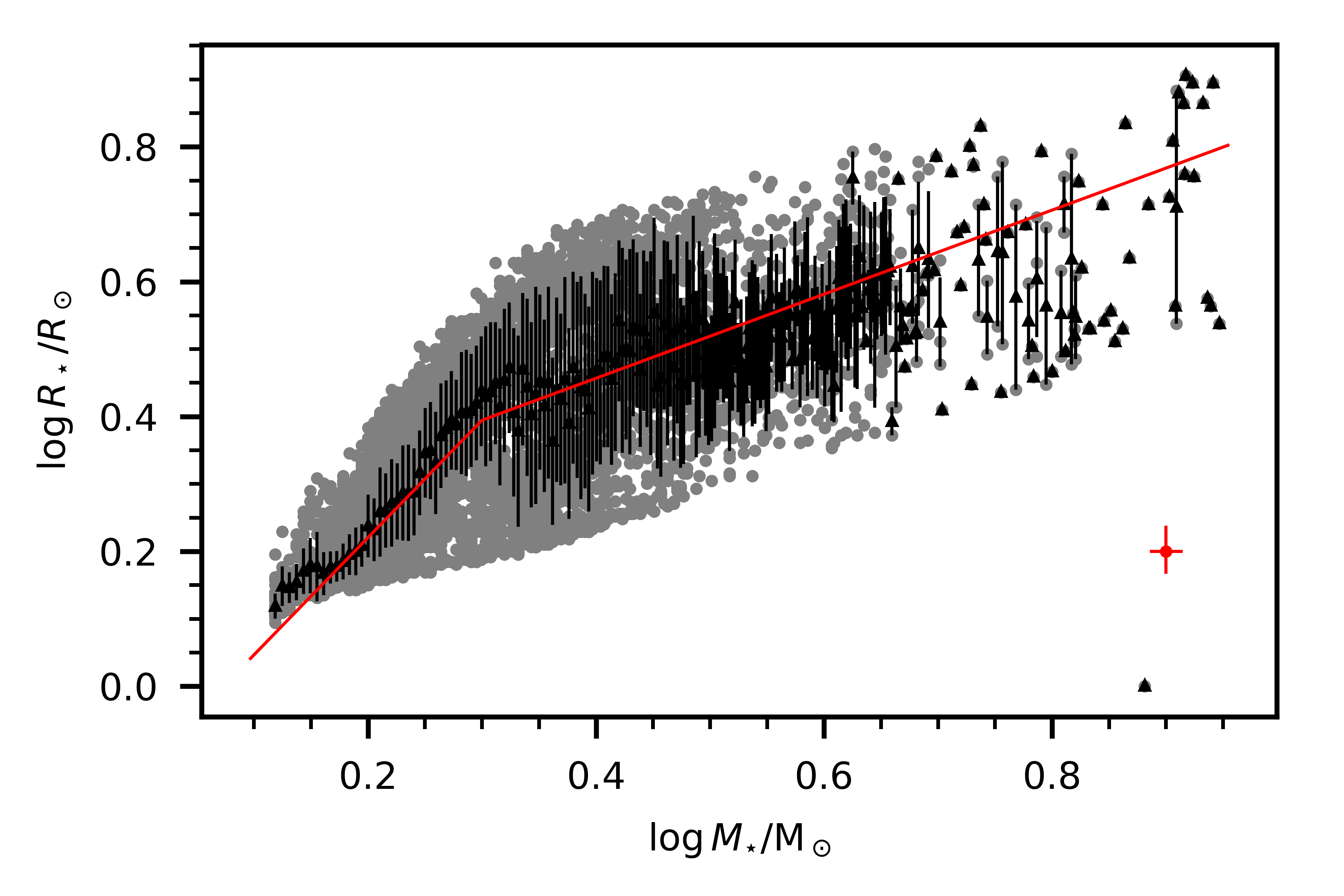}
    \caption{Same as Fig.~\ref{fig:mass_vs_radius}, but for $Z = 0.014$.}
    \label{fig:mass_vs_radius_Z140}
\end{figure}

\begin{figure}[htb]
    \centering
    \includegraphics[width = \columnwidth]{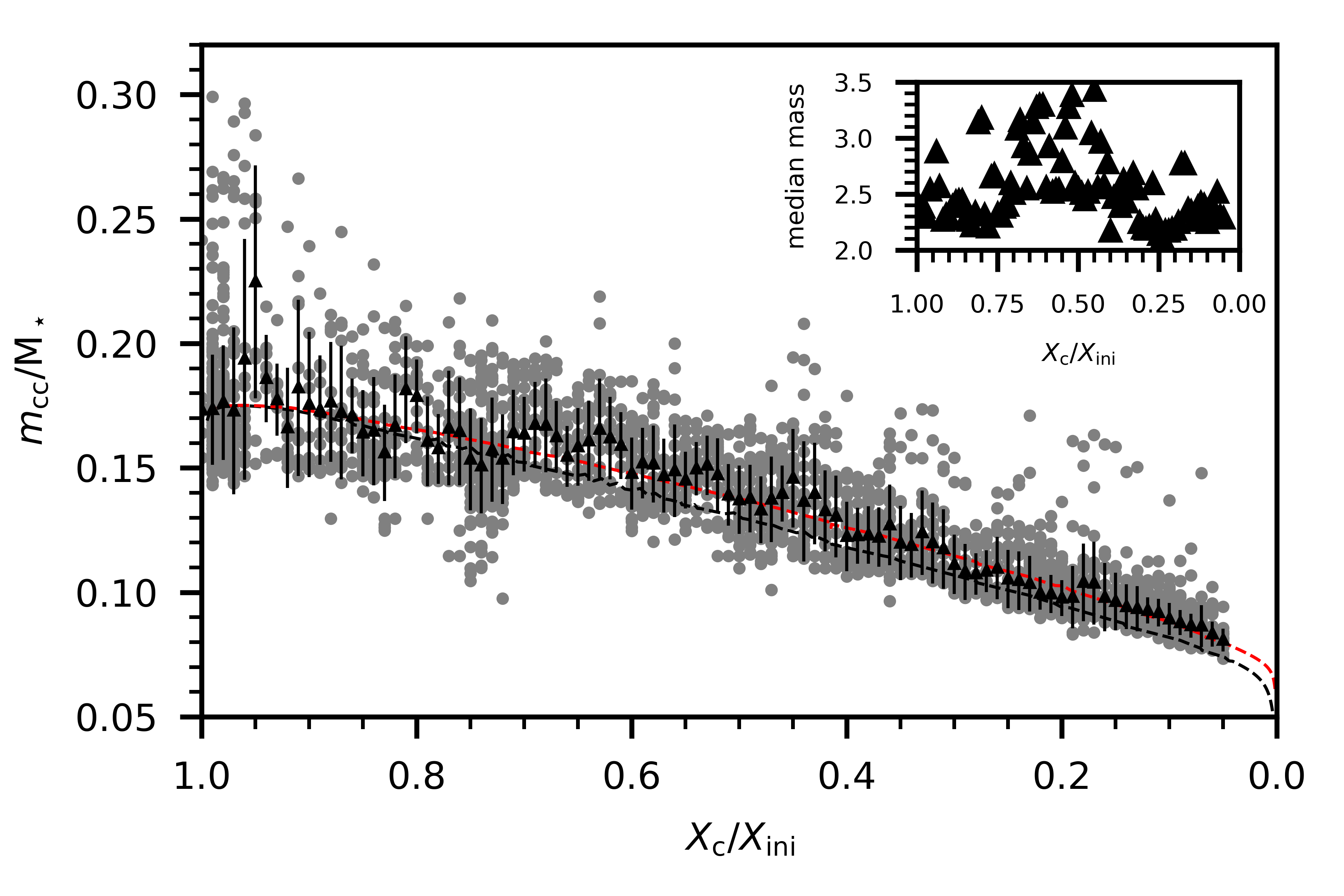}
    \caption{Same as Fig.~\ref{fig:xcx_vs_mcc}, but for $Z = 0.014$.}
    \label{fig:xcx_vs_mcc_Z140}
\end{figure}

\section{Accuracy of the methodology} \label{ap:method}
\begin{figure*}[htb]
    \centering
    \includegraphics[width = \textwidth]{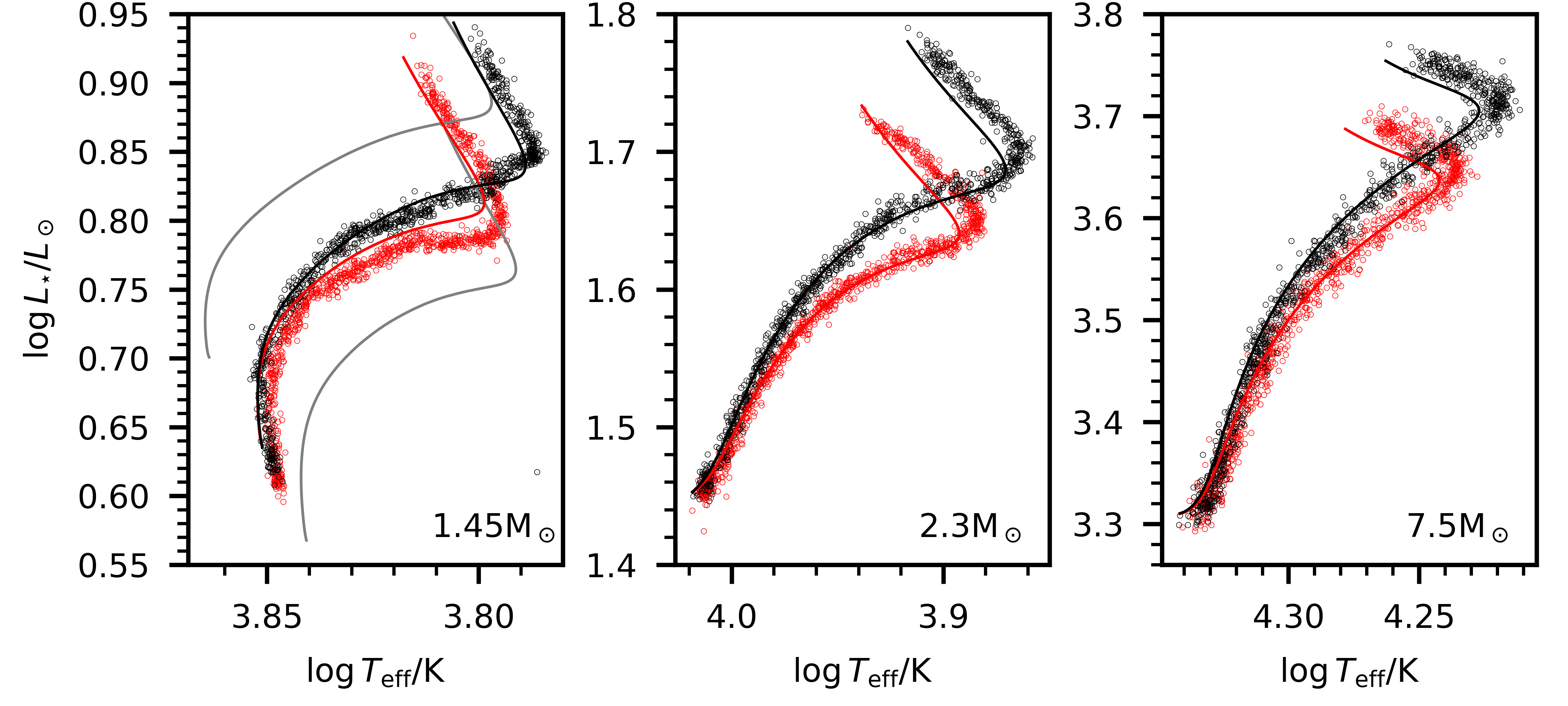}
    \caption{HRD showing stellar evolution tracks unseen by the CNF, and the sampling of these tracks by the CNF. The red colour indicates $f_{\rm CBM} = 0.01$, the black colour indicates $f_{\rm CBM} = 0.02$. In the left panel, tracks for 1.4 and 1.5\Msun ($f_{\rm CBM} = 0.015, \omega_0 = 0.05$) are shown in grey to illustrate the distance between two evolution tracks in the grid and the interpolation error of the CNF. }
    \label{fig:CNF_interpolation}
\end{figure*}

\edit{In this appendix, we discuss the accuracy of the methodology presented in this work. As a first test, we computed additional stellar evolution models that have not been used to train the CNF. Figure~\ref{fig:CNF_interpolation} shows the points sampled from the CNF compared to the actual models. The errors on the interpolated models are small enough that we would have to at least increase the number of masses by a factor five and similar for values of \fov (see left panel of Fig.~\ref{fig:CNF_interpolation} where the distance between two evolution models of different masses in grid are shown). Therefore, the relatively small error introduced by using a CNF is worth it.} 

\begin{figure*}[htb]
    \centering
    \includegraphics[width = \textwidth]{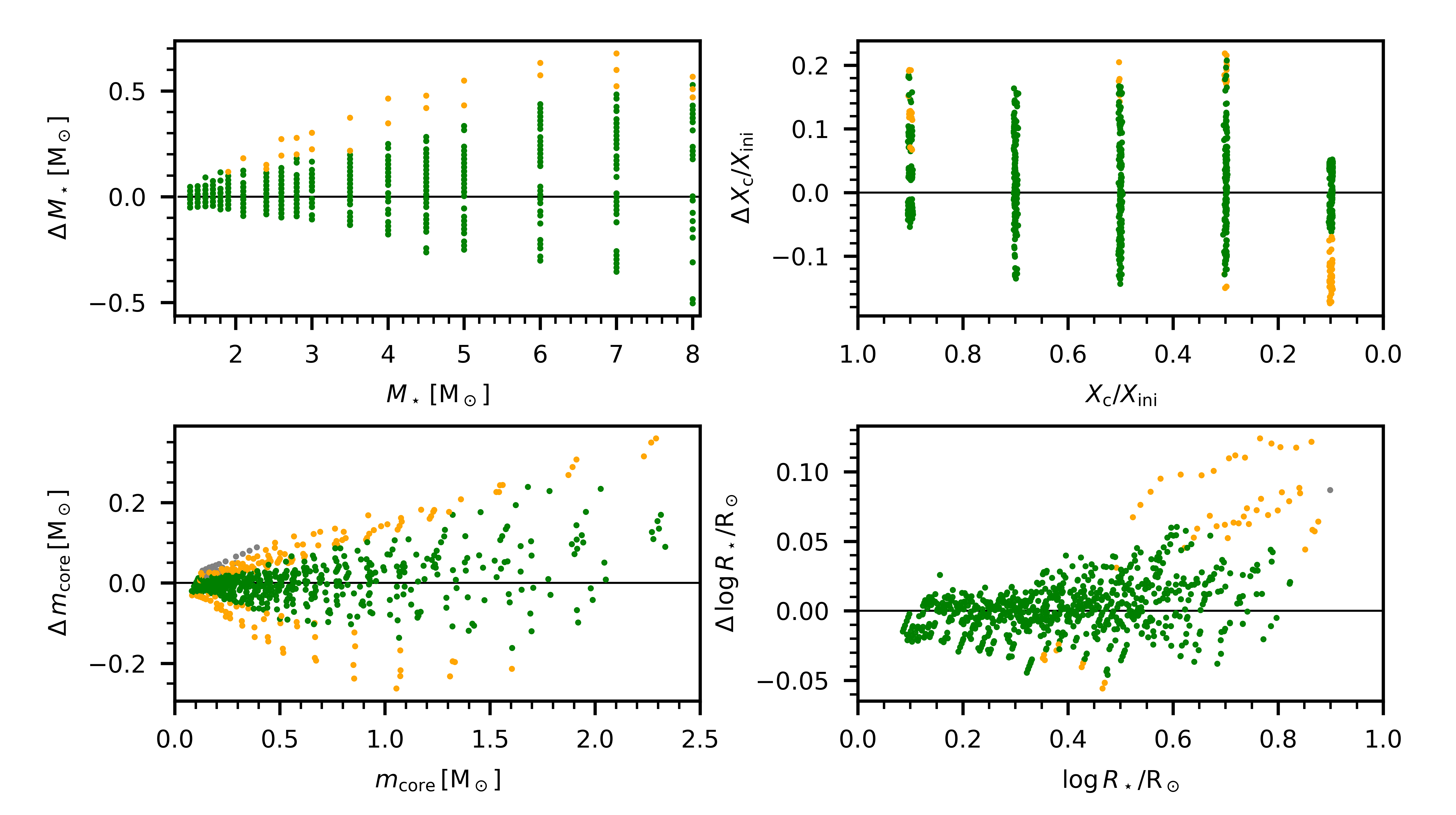}
    \caption{Discrepancy between the actual value and the recovered parameter for a set of validation data points ($\Delta x = x_{\rm actual} - x_{\rm recovered}$). The green colour indicates values that are recovered within the 68\%-confidence interval, the orange colour within twice this interval. }
    \label{fig:mock_data}
\end{figure*}

\edit{As a second test, we sampled points on the \mesa models and tried to recover the input parameters from the effective temperature and luminosity with the methodology described in Section~\ref{sect:method}. We assigned an uncertainty of 8\percent on $\log L_\star$ and 0.3\percent on $\log T_{\rm eff}$, representative of the typical uncertainties of the {\it Gaia\/} sample. Figure~\ref{fig:mock_data} shows the difference in the recovered parameters compared to the input value. Assuming these uncertainties, we recover in $\sim 75$\percent of the cases the actual convective core mass within the 68\%-confidence interval, while for the other parameters in more than 90\percent of the cases. In general, we do not observe any systematic offset in the discrepancies between the recovered values and actual values. Therefore, since we use a large sample of stars, these errors are mostly averaged out.} 

\begin{figure*}[htb]
    \centering
    \includegraphics[width = 0.8\textwidth]{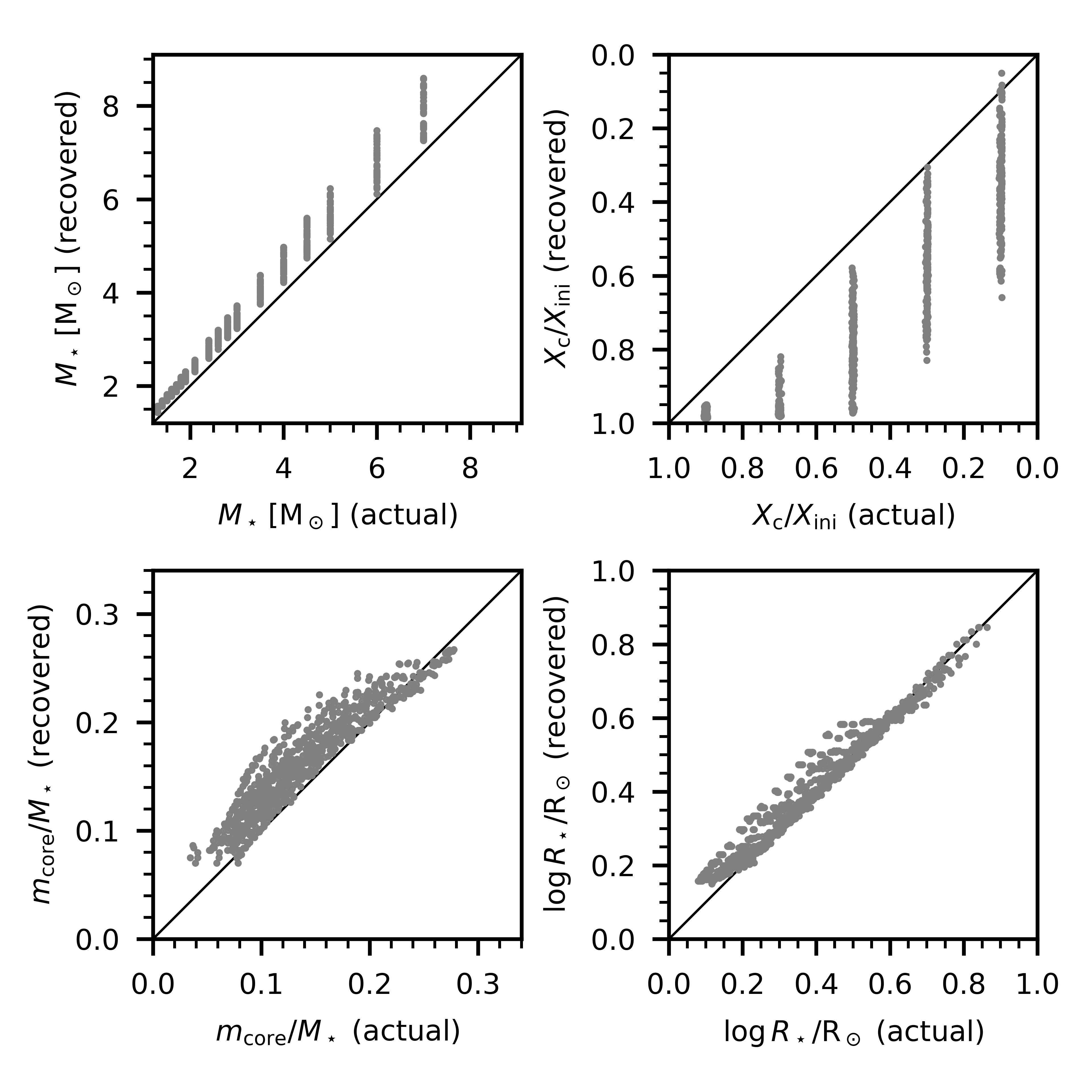}
    \caption{Actual value of the parameter in the mock data for $Z = 0.0045$ versus the recovered value when a metallicity of $Z= 0.014$ is assumed in the modelling. The solid black lines indicate the 1:1 line. }
    \label{fig:metallicity_test}
\end{figure*}

\edit{For the third test, we quantified the impact of assuming an incorrect metallicity in the modelling. We did this test for the most extreme case, that is, we modelled points from the $Z = 0.0045$ grid with the CNF that is trained on the $Z = 0.014$ grid. We note that we expect the majority of stars in the sample to have a metallicity between these two values \citep{deLaverny2024}. Figure~\ref{fig:metallicity_test} shows the recovered stellar parameters. The same uncertainties as previously mentioned are assumed for the luminosity and the effective temperature. The recovered mass is on average about 15\percent larger when an incorrect metallicity is assumed. For the fractional core masses, we find an average offset of roughly 0.02, as assuming a higher metallicity than in reality leads to finding younger stars. In general, the recovered radius is still correct, with some models for which the radius is overestimated.     }

\end{appendix}
\end{document}